\documentclass[aps,twocolumn,prb,tightenlines,floatfix,showpacs]{revtex4-1}
\usepackage{amssymb}
\usepackage{amsmath}
\usepackage{mathrsfs}
\usepackage{graphicx}
\usepackage{array}

\newcommand{\bs}[1]{\boldsymbol{#1}}

\newcommand{\abs}[1]{\left| #1 \right|}
\newcommand{\avg}[1]{\left\langle #1 \right\rangle}

\newcommand{\half}[0]{\frac{1}{2}}

\newcommand{\ria}[0]{\rightarrow}

\newcommand{\mrm}[1]{{\mathrm{#1}}}

\newcommand{\meq}[1]{\begin{equation} #1 \end{equation}}
\newcommand{\mea}[1]{\begin{align} #1 \end{align}}
\begin{document}

\title{Quantum oscillations in non-Fermi liquids:\\ 
Implications for high-temperature superconductors}

\author{Peter Scherpelz}
\author{Yan He}
\author{K. Levin}
\affiliation{James Franck Institute and Department of Physics,
University of Chicago, Chicago, Illinois 60637, USA}

\date{\today}

\begin{abstract}
We address quantum oscillation experiments in high $T_c$ superconductors
and the evidence from these experiments for a pseudogap versus a Fermi
liquid phase at high magnetic fields. As a concrete alternative to a
Fermi liquid phase, the pseudogap state we consider derives from earlier
work within a Gor'kov-based Landau level approach. Here the normal state
pairing gap in the presence of high fields is spatially non-uniform,
incorporating small gap values. These, in addition to $d$-wave gap
nodes, are responsible for the persistence of quantum oscillations.
Important here are methodologies for distinguishing different scenarios.
To this end we examine the temperature dependence of the oscillations.
Detailed quantitative analysis of this temperature dependence
demonstrates that a high field pseudogap state in the cuprates may well
``masquerade'' as a Fermi liquid.
\end{abstract}

\maketitle

The surprising discovery of quantum oscillations in the underdoped cuprate
high-temperature
superconductors\cite{doiron-leyraud_2007,yelland_2008,bangura_2008} 
can potentially elucidate
the normal, non-superconducting state of these materials.  However, a number of
experiments seem to indicate the presence of phenomena that
may be associated with the application of high magnetic fields,
rather than with the intrinsic normal phase.
The oscillatory frequency is very
small, which suggests a small and possibly reconstructed Fermi
surface.\cite{millis_2007,yao_2011,vignolle_2011,sebastian_2012}
More recently, static or quasi-static charge density 
wave order in the high-field regime has been observed, 
which also suggests substantial differences
between the zero-field normal state and the high-field
state.\cite{wu_2011,ghiringhelli_2012,chang_2012,wu_2013}
Changes
in magnetization, as well as transport coefficients,
 have also been found at high fields.\cite{li_2010,leboeuf_2007,*leboeuf_2011}

It is notable
that detailed studies of the
temperature dependence of the oscillatory amplitude\cite{sebastian_2010}
report excellent agreement with a Fermi-Dirac dependence,
providing no sign
of the pseudogap state that has been observed at magnetic field $H=0$.
Adding to the complexity,
specific heat measurements suggest an overall
temperature dependence of $\sqrt{H}$ that is consistent with the presence of a
$d$-wave
pairing gap.
\cite{riggs_2011}
These observations have led to
theoretical proposals in which there are
Fermi liquid-like features, perhaps co-existing with a pseudogap.
Two notable scenarios both introduce a
low frequency, third peak into the spectral function, while maintaining the
two other peaks at finite $\omega$ which reflect a gap
structure. \cite{senthil_2009,banerjee_2013}  
However,
the question of whether the quantum
oscillation measurements \textit{require} this peak (or more broadly, Fermi
liquid-like behavior near the Fermi surface) is an open and very 
important one. The answer bears on the proper microscopic description of
the cuprates.

In this paper we address this issue more phenomenologically.  
We quantitatively compare a high field pseudogap scenario and
an alternative
``co-existing pseudogap and Fermi
liquid" approach with a strict Fermi liquid. We do so via the 
measured 
\cite{sebastian_2010}
temperature dependences of
the quantum oscillations which have been interpreted to strongly support
a Fermi liquid phase at high $H$.
Our high field pseudogap scenario was derived 
earlier using a Gor'kov-based
theory \cite{scherpelz_2013,scherpelz_2013b} and incorporating Landau level physics. 
Although some model systems do display
substantial deviations, discrepancies for the
parameter sets appropriate to the cuprates are not
large enough to be detected in existing experiments. 
Thus we conclude that a non-Fermi liquid phase supports quantum
oscillations with $T$ dependence presently indistinguishable from that 
of a Fermi
liquid.

\begin{figure}[tb]
\includegraphics[clip = true, trim = 0.125in 0in 0.05in 0in, scale =
.45]{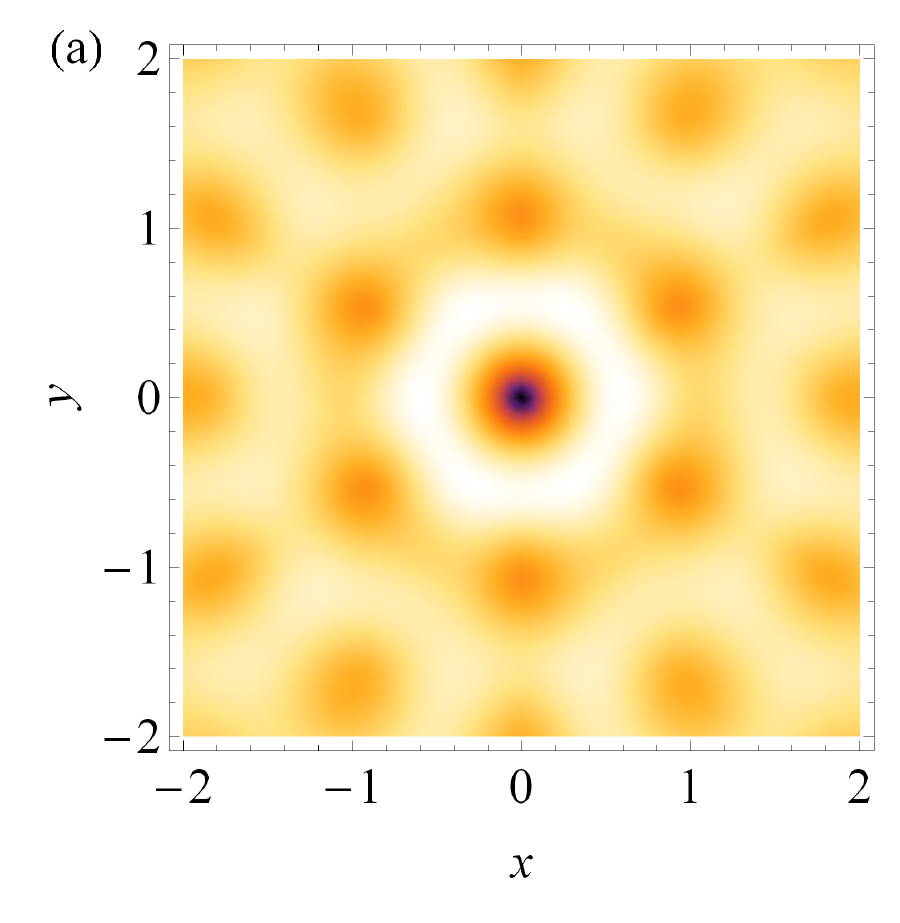}
\includegraphics[clip = true, trim = 0.125in 0in 0.05in 0in, scale =
.45]{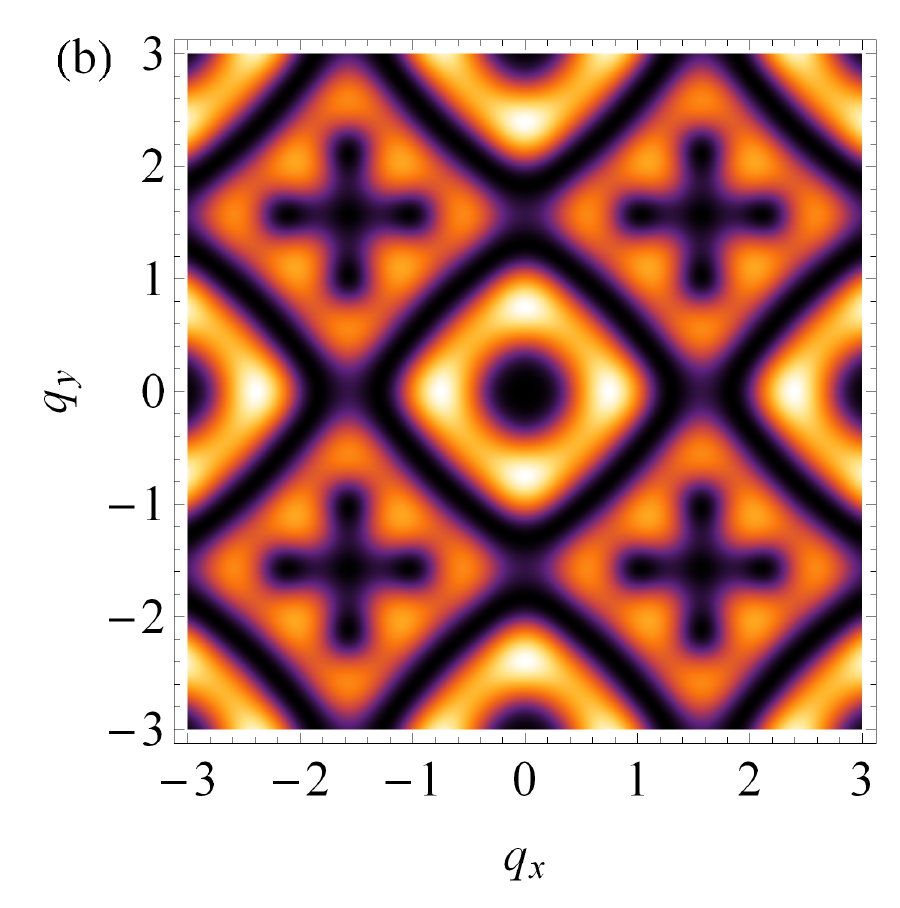}
\includegraphics[clip = true, trim = 0.3in -1in 0in 0in, scale =
.25]{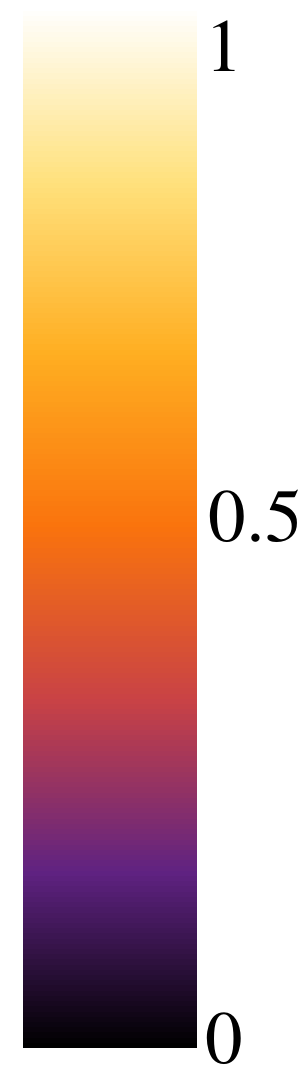}
\caption{\label{Fig0} \textbf{(a)} A density 
plot of $\abs{\Delta(\bs r)}^2$ for a 
pseudogap ``blurred vortex'' model.\cite{scherpelz_2013b}
This indicates that the normal state gap structure (representing
preformed pairs) is inhomogeneous
as a result of high magnetic fields.
\textbf{(b)} The density plot of $\abs{\Delta(\bs q)}^2$ for the $d$-wave, $n=8$
system used throughout the paper. 
Here $\bs q$ is
based on a magnetic translation group.
Both plots are normalized to
$\Delta_\mrm{max} = 1$.
}
\end{figure}

\textbf{Theory of the Pseudogap in High Magnetic Fields}
At $H=0$ the commonly used pseudogap self-energy
$\Sigma(\bs k,i\omega_n)$,
has been derived from a theory of pairing
fluctuations,\cite{janko_1997,maly_1997b} and has
also been obtained phenomenologically by fitting angle resolved photoemission
experiments.\cite{norman_1998,norman_2007}
It is given by
\meq{ \Sigma(\bs k,i\omega_n) 
= -i\gamma'+\frac{\Delta^2(\bs k)}{i\omega_n +\xi_{-\bs
k}+i\gamma}.\label{EqOurSE}}
where i$\omega_n$ is the fermionic Matsubara frequency, $\xi_{\bs k}$ the
single-particle dispersion, and $\gamma$ and $\gamma'$ damping coefficients 
associated with
the pairing gap 
($\Delta(\bs k)$)
and single particles, respectively.  

Using Gor'kov theory, we \cite{scherpelz_2013} and others \cite{dukan_1994} 
have shown earlier that in the presence of large
magnetic fields, with intra-Landau level pairing, the general BCS-like
structure of the Green's functions (and hence self energy as in
Eq.~\eqref{EqOurSE}) is maintained. 
For
this
quasi-two dimensional pseudogap state, we incorporate
Landau levels via $\xi(\bs k) \ria \xi(n,\bs q) =
(n+\half)\hbar\omega_c$, where $n$ is the Landau level and $\bs q$ the
degenerate quantum index. The latter is based on a magnetic translation group
 approach which is associated with the superconducting \cite{dukan_1994}
and pseudogap phases.\cite{scherpelz_2013}

Importantly,
a gap squared contribution (as in Eq.~\eqref{EqOurSE}) persists
\cite{scherpelz_2013}
into the normal phase.
This reflects pairing (as distinct from
phase) fluctuations which
arise from short-lived, preformed pairs; they
are to be associated with stronger-than-BCS attractive interactions
(consistent with high transition temperatures) and they lead to a pseudogap.
Furthermore, 
Gor'kov theory at high fields requires the
introduction of inhomogeneity in the gap function
$\Delta$.\cite{scherpelz_2013b}
Similarly, in the normal high-field state the pairing gap
$\Delta(\bs k)$ must be dependent on the $(n,\bs q)$ parameters defined
above, and this will lead to
real-space inhomogeneity.  Physically, these inhomogeneities reflect
excited pair states
which were shown \cite{scherpelz_2013b} to 
correspond to small distortions or excitations of the 
optimal (condensate) vortex configuration. As such, they represent
blurred lattice patterns.
This 
``precursor vortex''
state\cite{scherpelz_2013b} is illustrated in Fig.~\ref{Fig0}a.

For the purposes of this paper, the exact dependence of
$\Delta(n, \bs q)$
(where $\Delta$ is the (real) magnitude of the gap)
on $\bs q$ need not be determined. Instead
only the distribution of
$\Delta$ values over $\bs q$ is fixed.  
Throughout the paper we use the normalization that
$\avg{\abs{\Delta(\bs q)}^2} = \Delta$, the specified gap magnitude.
One can reasonably approximate
this
distribution by taking $\Delta(n,\bs q)$ independent
of $n$, as the distribution should change only slightly for moderate to large
Landau levels.  While the high field normal state gap inhomogeneity 
(``pseudovortex") is present for any
pairing symmetry,  it should be noted that
nodal effects from $d$-wave pairing in
a Landau level basis also lead to 
real-space inhomogeneity. 
\footnote{Note that in $s$-wave gaps, nodal states are created
solely by real-space inhomogeneity. However, in $d$-wave contributions
to nodal states from pairing symmetry and Landau level-based real-space
inhomogeneity are both present and become essentially
inseparable,\cite{vavilov_1998} which means we will not distinguish between
their descriptions here.}
For simplicity, we calculate the distribution for $n = 8$ from previous
work on $d$-wave pairing at high magnetic fields within the magnetic translation
group,\cite{vavilov_1998} 
and use this as a model distribution throughout the paper.  

\begin{figure}[tb]
\includegraphics[clip = true, trim = 1.25in 3.12in 4.5in 2.0in, scale =
.55]{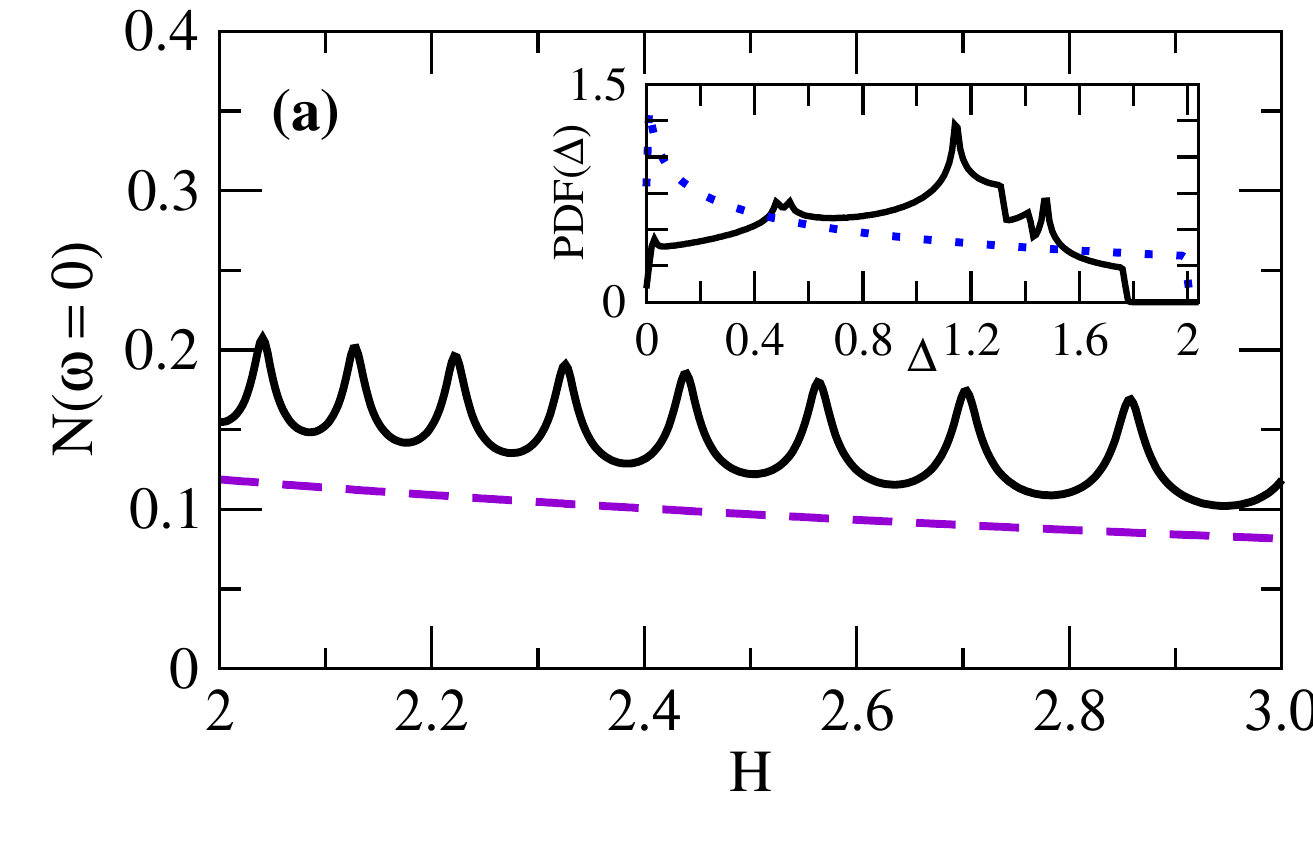}
\includegraphics[clip = true, trim = 1.25in 3.12in 4.5in 2.0in, scale =
.55]{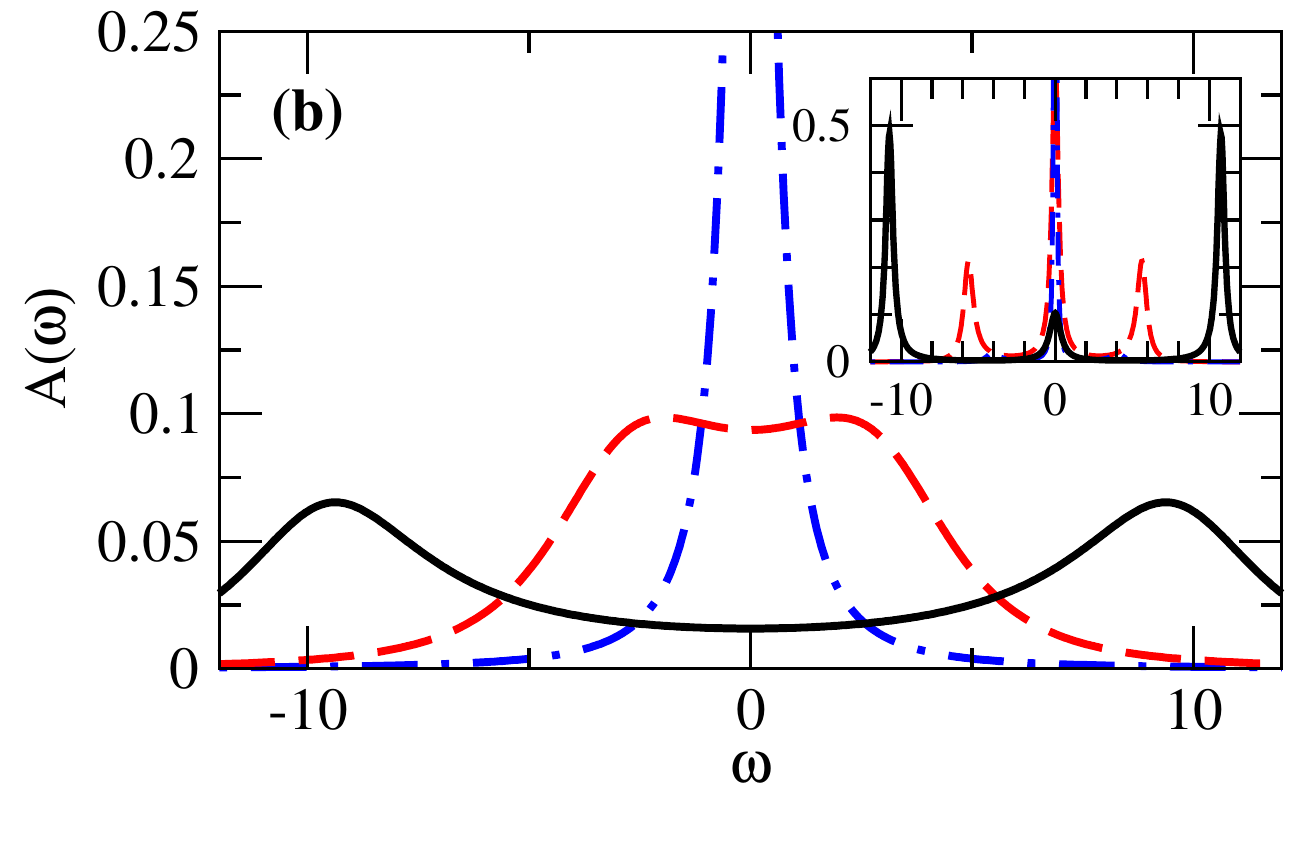}
\caption{\label{Fig1} \textbf{(a)} A plot of the density of states at the Fermi
surface vs.~$H$ (with $\hbar e / mc = 1$) for the 2D, high-field 
system with self-energy given
by Eq.~\eqref{EqOurSE}, with $\Delta = 10$, $\mu = 50$, $\gamma/\Delta = 0.5$
and $\gamma' = 0.2$. The black line shows the $d$-wave $\Delta(\bs q)$ used
throughout this paper, while the dashed purple line shows a constant $\Delta(\bs
q)$.  The inset shows the probability distribution function of $\Delta(\bs q)$
values (normalized so that $\sum_{\bs q} \Delta(\bs q)^2 = 1$) for the
high-field, $n = 8$ $d$-wave used here (black line) and for comparison a
distribution of
$\Delta(\bs k)$ for zero-field $d$-wave, $\Delta(\bs k) = \cos k_x - \cos k_y$
(blue dots).
\textbf{(b)} Example spectral functions for the system,
with variable $\Delta(\bs q)$.  In both the main plot and the inset, $\Delta
 = 10,4,1$ for the solid black, dashed red, and dash-dotted blue curve
respectively.  In the main plot, the self-energy in Eq.~\eqref{EqOurSE} is used
with $\gamma = 5$ and $\gamma' = 0.2$, while in the inset Eq.~(20) of
Ref.~\onlinecite{senthil_2009} is
used with $\Gamma = 4$ and $\gamma = 0.5$.}
\end{figure}

Fig.~\ref{Fig0}b
shows the density plot of $\abs{\Delta(\bs q)}^2$ while the Fig.~\ref{Fig1}a
inset shows the histogram used.  We
stress that deviations in vortex locations, pseudogap inhomogeneities, and
inter-Landau level pairing\cite{tesanovic_1998} 
should only change this distribution slightly, and
leave the conclusions unaffected.  Thus, 
with this method we have extended
the zero-field self energy, Eq.~\eqref{EqOurSE}, to a form appropriate
for the high-field pseudogap state.

It is useful to compare this high field normal state picture with others
in the literature, which focus on short-range fluctuations of the phase, both in
space\cite{stephen_1992} and time,\cite{banerjee_2013} caused by the disorder
introduced by moving vortices within a vortex liquid. 
By contrast, our theory does not assume the presence of
short-range phase coherence or vortices, but rather incorporates
fluctuating pair states.
We emphasize
Landau level physics that should occur in any paired, high-field system
and we focus
on nodal gap states
created by both the $d$-wave pairing symmetry and by real-space inhomogeneity.
Also important is the fact that
real-space inhomogeneity is necessary to allow a phase
transition (from the pseudogap to the superconducting state)
in a field, owing to the one dimensionality associated
with $H$. 
\cite{scherpelz_2013b} It is currently unclear whether true vortices
are present in the high-field quantum oscillation
regime, as different experiments arrive at different
conclusions.
\cite{li_2010,grissonnanche_2013,tan_2004}
In some ways, however, the vortex liquid theories and the present pseudogap
scenario are more similar than might be thought.
In the pseudogap scenario there is significant weight at small
gap values, while in the alternatives there is an additional
gapless (or Fermi liquid) spectral function peak.

\textbf{Quantum Oscillations in the Pseudogap State}
Using the self-energy, together with the form of $\Delta(\bs k)$, 
we address the amplitude of oscillations within
this non-Fermi liquid pseudogap state via calculations
of the density of
states at the Fermi surface, $N(\omega = 0)$. Here $N(\omega)
 = \sum_{\bs k} A(\bs
k,\omega),$ which will reflect the oscillations at zero-temperature.
The spectral function for the self-energy $A(n,\bs q,\omega) =
-2\mrm{Im }G(n,\bs q,\omega + i0)$ (see Fig.~\ref{Fig1}b).
Sample results for $N(0)$ vs.~$H$ are shown in Fig.~\ref{Fig1}a.\footnote{
We take $\gamma/\Delta$, which would be zero in the
superconducting state, to be as large as $0.5$ based on
estimates which use the Fermi arc size in 
zero-field.\cite{norman_2007,he_2013}  
The choice of $\gamma'$ has little effect on the conclusions, 
but
$\gamma' = 0.2$ strikes a balance between computability (with no extremely 
sharp features, as opposed to $\gamma' \ria 0$)
and the visibility of 
distinct oscillations (as opposed to large $\gamma'$). Parameters
used for the model from Ref.~\onlinecite{senthil_2009} preserve that paper's
choice of $\Gamma/\Delta$ and $\gamma/\Delta$, while $\Delta$ itself is scaled
to match the other models.}
While the
amplitude of the
oscillations ranges from 3 to 20 times smaller (depending on
$\gamma'$) than that  
in a Fermi liquid, 
they are still clearly visible.  Thus, a
non-Fermi liquid, pseudogap system can display robust quantum
oscillations, as we have shown in similar work near zero field.\cite{he_2013}
This is consistent with the quantum oscillations seen
previously within the superconducting phase of
extreme type-II 
superconductors. 
\cite{corcoran_1994,*maniv_1992,*norman_1995,*janssen_1998,
stephen_1992, dukan_1995}

A critical component of these oscillations is the presence of nodes or
near-nodal $\bs q$ states in $\Delta(\bs q)$, as Figs.~\ref{Fig0}b and
\ref{Fig1}a inset show are present in this system.  
If on the other hand one presumed
a constant $\Delta(\bs q)$,
Fig.~\ref{Fig1}a (which plots the
density of states) shows that
the amplitude of the oscillations is
 reduced to zero.  For the $d$-wave case
the near-nodal states, in contrast, preserve the
single-particle Landau level dispersion and thus the quantum oscillations.

\textbf{Temperature Dependence of Oscillations} \\
To distinguish a non-Fermi liquid from a
Fermi liquid 
on the basis of quantum oscillations, we next 
focus on the temperature dependence of the oscillation amplitude.
Importantly, this 
temperature dependence has previously been measured in YBa$_2$Cu$_3$O$_{6+x}$
and shown to have excellent agreement with Fermi liquid
theory.\cite{sebastian_2010} 
We compare the expected temperature dependence
of the high field pseudogap state developed here with that of 
an admixed Fermi liquid/ pseudogap scenario for the
normal state.\cite{senthil_2009} Throughout we presume
$\Delta(T)$ is roughly T-independent, since it depends on much
larger temperature scales than those accessed by quantum oscillations.

\begin{figure}[tb]
\includegraphics[clip = true, trim = 1.25in 3.12in 3.2in 1.2in, scale =
.50]{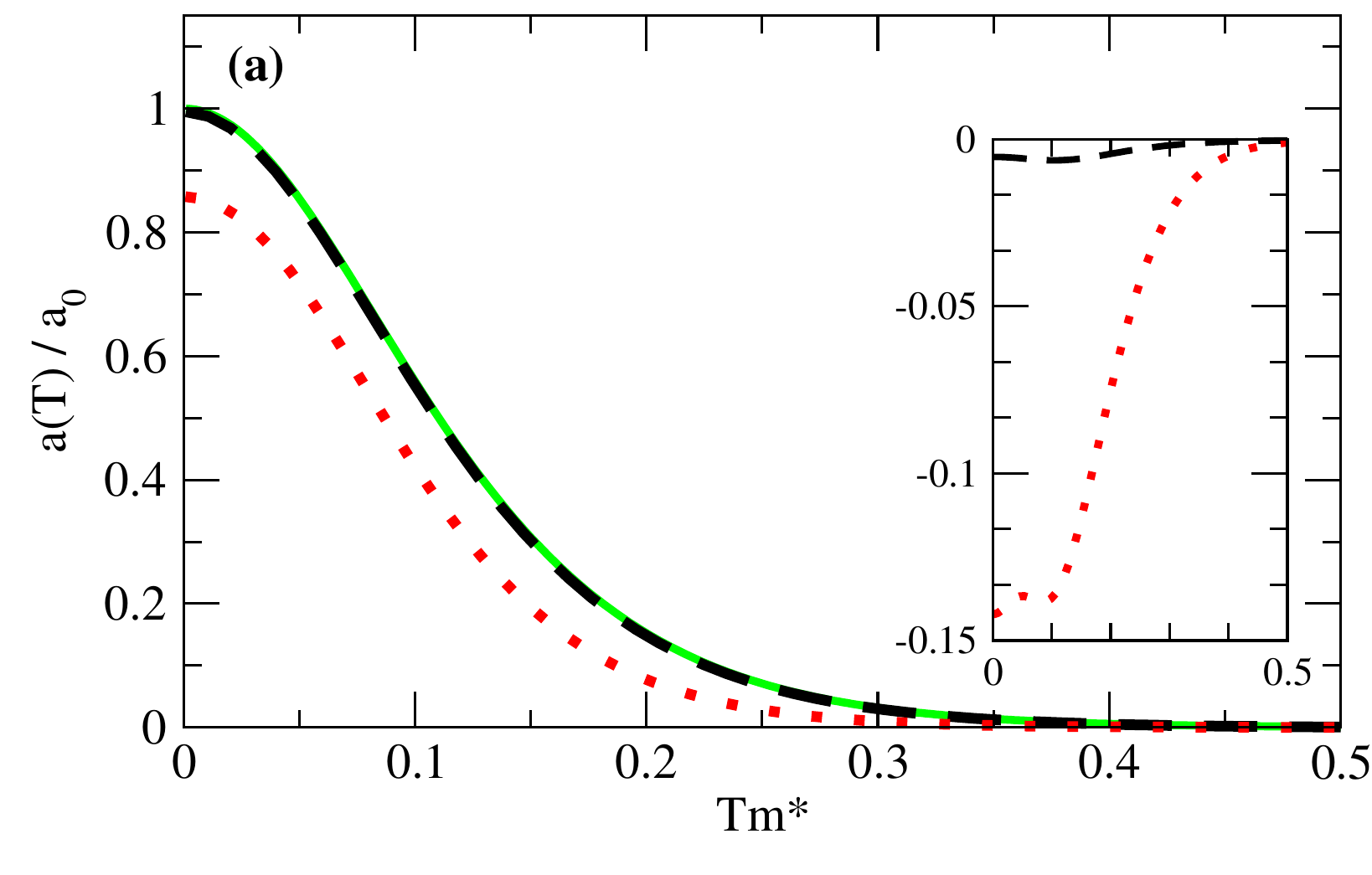}
\includegraphics[clip = true, trim = 1.25in 3.12in 3.2in 2.0in, scale =
.50]{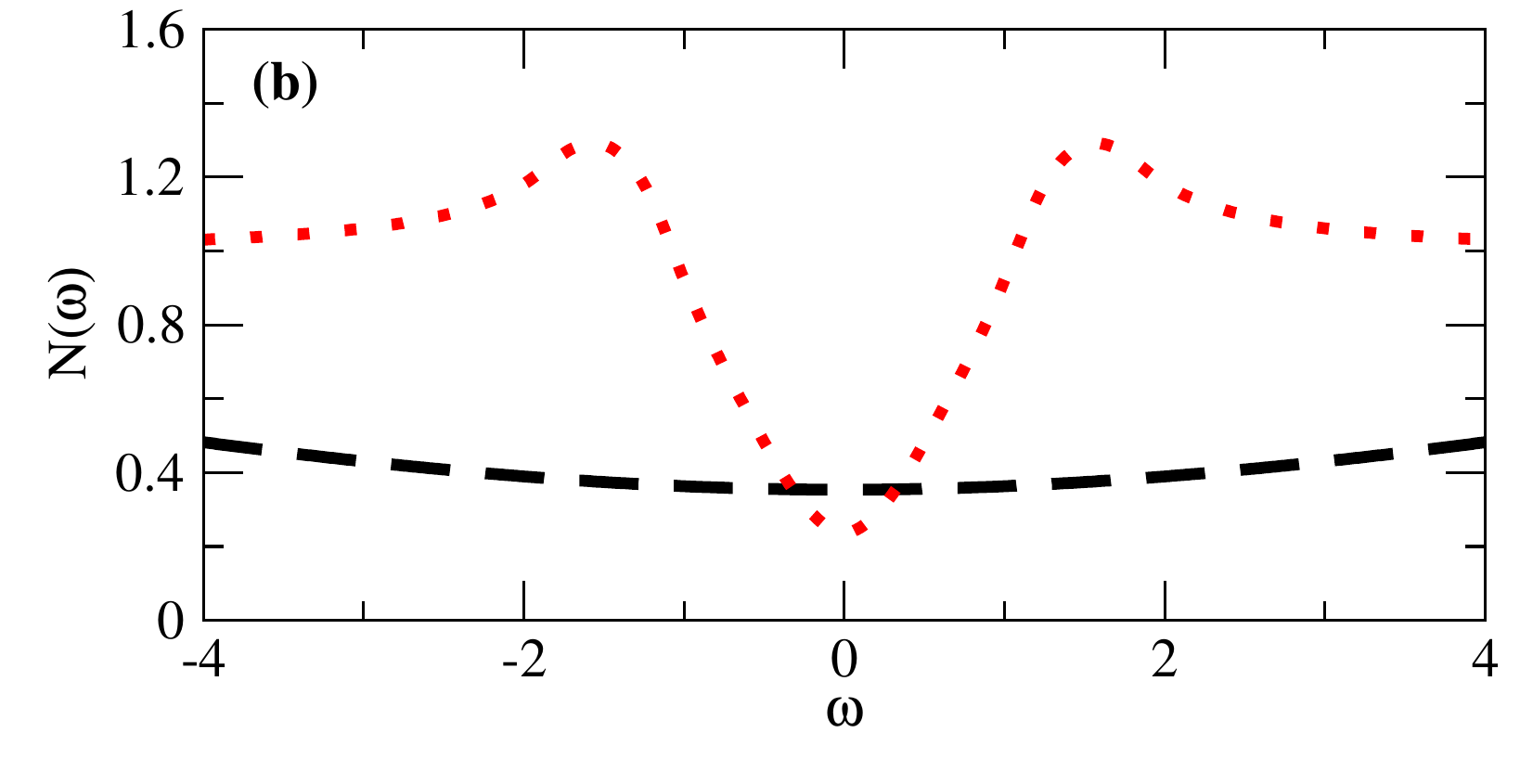}
\caption{\label{Fig2}\textbf{(a)} 
A plot of the rescaled temperature dependence of the oscillation
amplitude, $a(T)/a_0$ vs.~$Tm^*$.
The green solid line is for Fermi-liquid theory; the black dashed
line represents the present pseudogap model
with physical parameters $\Delta = 10\hbar\omega_c$ 
(Ref.~\onlinecite{shibauchi_2001}), 
$\gamma/\Delta = 0.5$ (consistent with Fermi 
arcs\cite{norman_2007,he_2013}), and 
$\gamma' = 0.2\hbar\omega_c$.  The red dotted line represents
an unphysical case for comparison, with small
$\Delta = \hbar\omega_c$ and $\gamma/\Delta = 0.1$.  The inset shows the
residual between the calculation and Fermi-liquid
theory.  \textbf{(b)} The non-oscillating density of states for the systems in
part (a).  The $\Delta = 10\hbar\omega_c$ 
case is quite flat
within the range $-1 \lesssim \omega \lesssim 1$.  In contrast, the
more unphysical $\Delta = \hbar\omega_c$ case 
shows a significant curvature of the density of states. The normalization
$\hbar\omega_c = 1$ is used here.}
\end{figure}

The temperature dependence is computed by investigating the total energy of
quasiparticles for a grand canonical system with fixed $\mu$:\cite{fetter_1971}
\meq{E(\mu,T) = \int_{-\infty}^\infty \frac{d\omega}{2\pi}
\sum_{\bs k} (\omega + \xi_{\bs k} + 2\mu)
 A(\bs k,\omega)f(\omega)}
 where $f(\omega) = (1 +
\exp(\omega/T))^{-1}$ is the Fermi function. Because the oscillations are
created by quasiparticles near the Fermi surface, 
$\omega$ and $\xi_{\bs k}$ are both
much less than $\mu$, and we can approximate $\omega + \xi_{\bs k} + 2\mu \ria
2\mu$. 

Following earlier work \cite{abrikosov_1988,
dukan_1995} we take the large Landau level limit and use the
Poisson resummation formula to extract the fundamental frequency of
oscillation. 
After integration by parts
this yields \cite{shoenberg_1984}
\mea{\tilde{E}(\mu,T) = &-2\mu\int_{-\infty}^\infty dy\frac{df(y)}{dy}
\int_{-\infty}^y \frac{d\omega}{2\pi}
\int_{-\infty}^\infty \frac{d\xi^0}{\hbar\omega_c}  \notag \\
&{}\times\sum_{\bs q}
e^{2\pi i \xi^0/(\hbar \omega_c)}
A(\xi^0-\mu,\bs q,\omega),\label{EqFinalTD} }
where $\xi^0 = \xi(n) + \mu$.
In a Fermi liquid system, in which $A$ only depends on $\xi - \omega$, this
becomes
a convolution
and gives a total amplitude equal
to the
Fourier transform of $\abs{df(y)/dy}$ as in previous 
work.\cite{shoenberg_1984,sebastian_2010}
(Note that throughout we take $A$ to be temperature independent.)
For a general 
non-Fermi liquid system, the above amplitude will not be a convolution, and the
temperature-dependent amplitude can instead be calculated using an analytic form
of $A$.\footnote{For our calculation, 
we make one more transformation, by splitting up the $\omega$ integral in
Eq.~\eqref{EqFinalTD} (with $I$
the integrand):
$\int_{-\infty}^y d\omega\ I(\omega) \ria \int_0^y d\omega\ I(\omega) +
\int_{-\infty}^0 d\omega\ I(\omega).$ We know that as $T \ria \infty$, the
result of Eq.~\eqref{EqFinalTD} must go to zero, and with $\int_{-\infty}^\infty
dy\ (df(y)/dy) = 1$, we find that $\int_{-\infty}^0 d\omega\ I(\omega) =
-\lim_{T \ria \infty} \int_{-\infty}^\infty dy\ (df(y)/dy) 
\int_{0}^y d\omega\ I(\omega)$.  This latter formula is used to compute the
constant part of the original integral, so that each individual calculation 
only requires
computation of the $\int_0^y d\omega\ I(\omega)$ term.}

\textbf{Results}
The calculated 
temperature dependence of this non-Fermi liquid with a large pseudogap 
is shown as the dashed line
in Fig.~\ref{Fig2}a and gives excellent agreement with Fermi liquid theory
(solid green line).
In contrast, the dotted curve with a much smaller $\Delta = \hbar\omega_c$
displays large deviations from the Fermi liquid.
In the cuprates
$\Delta$ is much larger than $\hbar\omega_c$.  (For example, in
Ref.~\onlinecite{sebastian_2010} $\hbar\omega_{c,\mrm{max}} = \hbar e 
B_\mrm{max} / m^* = 3.8$ meV, whereas $\Delta$ can be
 tens of meV.\cite{shibauchi_2001}) This indicates that
it can be possible for a pseudogapped system to
``masquerade'' as a Fermi liquid for quantum oscillation measurements in the
cuprates.

The
discrepancy in the energy scales of $\Delta$ (Ref.~\onlinecite{shibauchi_2001})
 and $\hbar\omega_c$ 
is the primary cause of this similarity to Fermi
liquid behavior. As shown by the dashed line in
Fig.~\ref{Fig2}b, because these scales are so different, the underlying
density of states has very little curvature.
This, as well as
the damping and inhomogeneity, 
creates a flat density of states near the Fermi surface which in turn
leads to Fermi liquid-like behavior in the oscillations.
Specifically, Fig.~\ref{Fig2}b
plots the non-oscillating integrand
$N(\omega) = \int_{-\infty}^\infty d\xi^0/(\hbar\omega_c) 
\sum_{\bs q}
A(\xi^0-\mu,\bs q,\omega)$, 
essentially a zero-field density of states.  As quantum
oscillations are primarily sensitive to the innermost Landau level to the Fermi
surface, Eq.~\eqref{EqFinalTD} is mostly sensitive to effects within  
$\hbar\omega_c$ of the Fermi surface.  Although systems with $\Delta =
\hbar\omega_c$ do display variations on this scale, in more
physical situations
as studied here 
they are negligible.

\begin{table}[tb]
\begin{tabular}{lccc}
Moment $\mu_K$ & $\mu_6$ & $\mu_8$ & $\mu_{10}$ \\ \hline
FL Theory & $874.8$ & $2.697 \times 10^4$ & $9.937 \times 10^5$ \\
Ref.~\onlinecite{sebastian_2010} & $873$ & $2.65\times 10^4$ 
& $9.49 \times 10^5$ \\
(1) $\Delta = 10\hbar\omega_c$ & $872.6$ & $2.682 \times 10^4$ & $9.859 \times
10^5$ \\ % From 20130917D10_b
(2) $\Delta = \hbar\omega_c$ & $797.1$ & $2.163 \times 10^4$ & $6.997 
\times 10^5$ 
\\ % From 20130919D1_a
(3) Ref.~\onlinecite{senthil_2009} & $875.0$ & $2.698 \times 10^4$ & $9.948
\times 10^5$ % From AbsTDs1t3THRPKD10New from 20130919a 
%All updated to current version
\end{tabular}
\caption{\label{Tab1} A table of $z$-limited moments $\mu_K = 
\int_{-7.5}^{7.5} z^K
\abs{f'(z)} dz$ for different cases.   (1)
and (2) use the parameters from the dashed black line and dotted red line of
Fig.~\ref{Fig1}, respectively.  (3) uses the spectral function in
Ref.~\onlinecite{senthil_2009} with $\Delta = 10\hbar\omega_c$, $\Gamma = 4$ and
$\gamma = 0.5$.  Similarly to Ref.~\onlinecite{sebastian_2010} 
we fit the zero-temperature amplitude $a_0$
and the effective mass $m^*$ to the first two even 
moments which are obtained exactly.  The next three
even moments are non-fitted results, presented here.
(Note that
Ref.~\onlinecite{sebastian_2010} partially uses these higher
 moments in their fit.)}
\end{table}

In order to more quantitatively determine the quality of fit, 
we calculate the moments of the Fourier transform of the amplitude, as in
Ref.~\onlinecite{sebastian_2010}, which are displayed in 
Table~\ref{Tab1}.\footnote{Note that unlike Eq.~(1) in
Ref.~\onlinecite{sebastian_2010},
 no $\eta_\mrm{lim}$ is used
here.}
In the table, 
Case (1) uses the large $\Delta = 10
\hbar\omega_c$
and $\gamma = 0.5\Delta$ considered throughout the paper, 
and shows excellent agreement with the
theoretical Fermi liquid values.  Case (2) provides an example of a non-Fermi
liquid that does display large deviations from a Fermi liquid system, with a
small $\Delta = \hbar\omega_c$, as can clearly be seen in Fig.~\ref{Fig2}a as
well.  This provides evidence that this technique can be useful for the
discrimination of non-Fermi liquids in some systems.
Finally,
in Case (3) we consider a different
theoretical proposal, using the self-energy in Eq.~(20) of
Ref.~\onlinecite{senthil_2009}.  This system produces a three-peaked spectral
function as shown in the inset of Fig.~\ref{Fig1}b, which provides more weight 
at the Fermi surface in order to restore Fermi liquid behavior.
\cite{senthil_2009}
Excellent agreement with Fermi liquid theory is
obtained. This demonstrates
that in the non-Fermi liquid model of Case (1)
deviation from Fermi liquid behavior is present; however, it is
not 
detectable with existing experiments.

\textbf{Conclusions} 
It should be stressed that the goal of the present paper was not
to introduce a scenario for a high field non-Fermi liquid
normal state (which was presented earlier in other
contexts
\cite{scherpelz_2013,scherpelz_2013b}).
This is
admittedly a very controversial and still unresolved subject.
Rather the aim of this paper
is to design and present tests of different existing (and possibly future)
theoretical scenarios by
addressing the magnitude and temperature dependence of observed
quantum oscillations.

By starting with a specific, non-Fermi liquid model of a high-field pseudogap, 
we have found that quantum oscillations can both be present in a non-Fermi
liquid and display
a temperature dependence remarkably similar to that of a Fermi liquid.  
Here we
have focused on a model, built on Gor'kov theory,
which incorporates
Landau-level based pairing, real-space
inhomogeneity, and $d$-wave pairing symmetry.
Our non-Fermi liquid scenario is to be contrasted with
hybrid Fermi liquid-pseudogap approaches which view the ``normal"
high field phase as a vortex liquid, \cite{senthil_2009,banerjee_2013} 
a concept about which there is
not yet unanimity.
\cite{li_2010,grissonnanche_2013}
Within our model, two major components are responsible for the robust
oscillations. First, the nodal or
near-nodal effects leading to small
values of $\Delta(\bs q)$ are critical in retaining the visibility of
quantum oscillations in these systems.  Second, the large discrepancy in energy
scales between the gap\cite{shibauchi_2001} 
and the cyclotron frequency makes the system
appear relatively
Fermi liquid-like on the scale of the oscillations.

The fact that existing experimental work has
not distinguished between these scenarios
may allow for a simple resolution between oscillatory measurements and
the specific heat measurements which suggest the continued presence
of a $d$-wave gap at high fields.\cite{riggs_2011}
We note that the temperature dependent formalism outlined in
this paper may serve as a vehicle for testing future theories of these
oscillatory phenomena.
Additionally, our work has shown that an extremely high level of precision will
be required in future experiments to distinguish among
different theoretical
scenarios of the oscillations.

This work is supported by NSF-MRSEC Grant 0820054. P.S.~acknowledges 
support from the Hertz Foundation.

\bibliography{Review3}

\ifx\undefined\allcaps\def\allcaps#1{#1}\fi
\begin{thebibliography}{44}%
\makeatletter
\providecommand \@ifxundefined [1]{%
 \@ifx{#1\undefined}
}%
\providecommand \@ifnum [1]{%
 \ifnum #1\expandafter \@firstoftwo
 \else \expandafter \@secondoftwo
 \fi
}%
\providecommand \@ifx [1]{%
 \ifx #1\expandafter \@firstoftwo
 \else \expandafter \@secondoftwo
 \fi
}%
\providecommand \natexlab [1]{#1}%
\providecommand \enquote  [1]{``#1''}%
\providecommand \bibnamefont  [1]{#1}%
\providecommand \bibfnamefont [1]{#1}%
\providecommand \citenamefont [1]{#1}%
\providecommand \href@noop [0]{\@secondoftwo}%
\providecommand \href [0]{\begingroup \@sanitize@url \@href}%
\providecommand \@href[1]{\@@startlink{#1}\@@href}%
\providecommand \@@href[1]{\endgroup#1\@@endlink}%
\providecommand \@sanitize@url [0]{\catcode `\\12\catcode `\$12\catcode
  `\&12\catcode `\#12\catcode `\^12\catcode `\_12\catcode `\%12\relax}%
\providecommand \@@startlink[1]{}%
\providecommand \@@endlink[0]{}%
\providecommand \url  [0]{\begingroup\@sanitize@url \@url }%
\providecommand \@url [1]{\endgroup\@href {#1}{\urlprefix }}%
\providecommand \urlprefix  [0]{URL }%
\providecommand \Eprint [0]{\href }%
\providecommand \doibase [0]{http://dx.doi.org/}%
\providecommand \selectlanguage [0]{\@gobble}%
\providecommand \bibinfo  [0]{\@secondoftwo}%
\providecommand \bibfield  [0]{\@secondoftwo}%
\providecommand \translation [1]{[#1]}%
\providecommand \BibitemOpen [0]{}%
\providecommand \bibitemStop [0]{}%
\providecommand \bibitemNoStop [0]{.\EOS\space}%
\providecommand \EOS [0]{\spacefactor3000\relax}%
\providecommand \BibitemShut  [1]{\csname bibitem#1\endcsname}%
\let\auto@bib@innerbib\@empty
%</preamble>
\bibitem [{\citenamefont {Doiron-Leyraud}\ \emph {et~al.}(2007)\citenamefont
  {Doiron-Leyraud}, \citenamefont {Proust}, \citenamefont {{LeBoeuf}},
  \citenamefont {Levallois}, \citenamefont {Bonnemaison}, \citenamefont
  {Liang}, \citenamefont {Bonn}, \citenamefont {Hardy},\ and\ \citenamefont
  {Taillefer}}]{doiron-leyraud_2007}%
  \BibitemOpen
  \bibfield  {author} {\bibinfo {author} {\bibfnamefont {N.}~\bibnamefont
  {Doiron-Leyraud}}, \bibinfo {author} {\bibfnamefont {C.}~\bibnamefont
  {Proust}}, \bibinfo {author} {\bibfnamefont {D.}~\bibnamefont {{LeBoeuf}}},
  \bibinfo {author} {\bibfnamefont {J.}~\bibnamefont {Levallois}}, \bibinfo
  {author} {\bibfnamefont {J.-B.}\ \bibnamefont {Bonnemaison}}, \bibinfo
  {author} {\bibfnamefont {R.}~\bibnamefont {Liang}}, \bibinfo {author}
  {\bibfnamefont {D.~A.}\ \bibnamefont {Bonn}}, \bibinfo {author}
  {\bibfnamefont {W.~N.}\ \bibnamefont {Hardy}}, \ and\ \bibinfo {author}
  {\bibfnamefont {L.}~\bibnamefont {Taillefer}},\ }\href {\doibase
  10.1038/nature05872} {\bibfield  {journal} {\bibinfo  {journal} {Nature}\
  }\textbf {\bibinfo {volume} {447}},\ \bibinfo {pages} {565} (\bibinfo {year}
  {2007})}\BibitemShut {NoStop}%
\bibitem [{\citenamefont {Yelland}\ \emph {et~al.}(2008)\citenamefont
  {Yelland}, \citenamefont {Singleton}, \citenamefont {Mielke}, \citenamefont
  {Harrison}, \citenamefont {Balakirev}, \citenamefont {Dabrowski},\ and\
  \citenamefont {Cooper}}]{yelland_2008}%
  \BibitemOpen
  \bibfield  {author} {\bibinfo {author} {\bibfnamefont {E.~A.}\ \bibnamefont
  {Yelland}}, \bibinfo {author} {\bibfnamefont {J.}~\bibnamefont {Singleton}},
  \bibinfo {author} {\bibfnamefont {C.~H.}\ \bibnamefont {Mielke}}, \bibinfo
  {author} {\bibfnamefont {N.}~\bibnamefont {Harrison}}, \bibinfo {author}
  {\bibfnamefont {F.~F.}\ \bibnamefont {Balakirev}}, \bibinfo {author}
  {\bibfnamefont {B.}~\bibnamefont {Dabrowski}}, \ and\ \bibinfo {author}
  {\bibfnamefont {J.~R.}\ \bibnamefont {Cooper}},\ }\href {\doibase
  10.1103/PhysRevLett.100.047003} {\bibfield  {journal} {\bibinfo  {journal}
  {Phys. Rev. Lett.}\ }\textbf {\bibinfo {volume} {100}},\ \bibinfo {pages}
  {047003} (\bibinfo {year} {2008})}\BibitemShut {NoStop}%
\bibitem [{\citenamefont {Bangura}\ \emph {et~al.}(2008)\citenamefont
  {Bangura}, \citenamefont {Fletcher}, \citenamefont {Carrington},
  \citenamefont {Levallois}, \citenamefont {Nardone}, \citenamefont {Vignolle},
  \citenamefont {Heard}, \citenamefont {Doiron-Leyraud}, \citenamefont
  {{LeBoeuf}}, \citenamefont {Taillefer}, \citenamefont {Adachi}, \citenamefont
  {Proust},\ and\ \citenamefont {Hussey}}]{bangura_2008}%
  \BibitemOpen
  \bibfield  {author} {\bibinfo {author} {\bibfnamefont {A.~F.}\ \bibnamefont
  {Bangura}}, \bibinfo {author} {\bibfnamefont {J.~D.}\ \bibnamefont
  {Fletcher}}, \bibinfo {author} {\bibfnamefont {A.}~\bibnamefont
  {Carrington}}, \bibinfo {author} {\bibfnamefont {J.}~\bibnamefont
  {Levallois}}, \bibinfo {author} {\bibfnamefont {M.}~\bibnamefont {Nardone}},
  \bibinfo {author} {\bibfnamefont {B.}~\bibnamefont {Vignolle}}, \bibinfo
  {author} {\bibfnamefont {P.~J.}\ \bibnamefont {Heard}}, \bibinfo {author}
  {\bibfnamefont {N.}~\bibnamefont {Doiron-Leyraud}}, \bibinfo {author}
  {\bibfnamefont {D.}~\bibnamefont {{LeBoeuf}}}, \bibinfo {author}
  {\bibfnamefont {L.}~\bibnamefont {Taillefer}}, \bibinfo {author}
  {\bibfnamefont {S.}~\bibnamefont {Adachi}}, \bibinfo {author} {\bibfnamefont
  {C.}~\bibnamefont {Proust}}, \ and\ \bibinfo {author} {\bibfnamefont {N.~E.}\
  \bibnamefont {Hussey}},\ }\href {\doibase 10.1103/PhysRevLett.100.047004}
  {\bibfield  {journal} {\bibinfo  {journal} {Phys. Rev. Lett.}\ }\textbf
  {\bibinfo {volume} {100}},\ \bibinfo {pages} {047004} (\bibinfo {year}
  {2008})}\BibitemShut {NoStop}%
\bibitem [{\citenamefont {Millis}\ and\ \citenamefont
  {Norman}(2007)}]{millis_2007}%
  \BibitemOpen
  \bibfield  {author} {\bibinfo {author} {\bibfnamefont {A.~J.}\ \bibnamefont
  {Millis}}\ and\ \bibinfo {author} {\bibfnamefont {M.~R.}\ \bibnamefont
  {Norman}},\ }\href {\doibase 10.1103/PhysRevB.76.220503} {\bibfield
  {journal} {\bibinfo  {journal} {Phys. Rev. B}\ }\textbf {\bibinfo {volume}
  {76}},\ \bibinfo {pages} {220503} (\bibinfo {year} {2007})}\BibitemShut
  {NoStop}%
\bibitem [{\citenamefont {Yao}\ \emph {et~al.}(2011)\citenamefont {Yao},
  \citenamefont {Lee},\ and\ \citenamefont {Kivelson}}]{yao_2011}%
  \BibitemOpen
  \bibfield  {author} {\bibinfo {author} {\bibfnamefont {H.}~\bibnamefont
  {Yao}}, \bibinfo {author} {\bibfnamefont {D.-H.}\ \bibnamefont {Lee}}, \ and\
  \bibinfo {author} {\bibfnamefont {S.}~\bibnamefont {Kivelson}},\ }\href
  {\doibase 10.1103/PhysRevB.84.012507} {\bibfield  {journal} {\bibinfo
  {journal} {Phys. Rev. B}\ }\textbf {\bibinfo {volume} {84}},\ \bibinfo
  {pages} {012507} (\bibinfo {year} {2011})}\BibitemShut {NoStop}%
\bibitem [{\citenamefont {Vignolle}\ \emph {et~al.}(2011)\citenamefont
  {Vignolle}, \citenamefont {Vignolles}, \citenamefont {{LeBoeuf}},
  \citenamefont {Lepault}, \citenamefont {Ramshaw}, \citenamefont {Liang},
  \citenamefont {Bonn}, \citenamefont {Hardy}, \citenamefont {Doiron-Leyraud},
  \citenamefont {Carrington}, \citenamefont {Hussey}, \citenamefont
  {Taillefer},\ and\ \citenamefont {Proust}}]{vignolle_2011}%
  \BibitemOpen
  \bibfield  {author} {\bibinfo {author} {\bibfnamefont {B.}~\bibnamefont
  {Vignolle}}, \bibinfo {author} {\bibfnamefont {D.}~\bibnamefont {Vignolles}},
  \bibinfo {author} {\bibfnamefont {D.}~\bibnamefont {{LeBoeuf}}}, \bibinfo
  {author} {\bibfnamefont {S.}~\bibnamefont {Lepault}}, \bibinfo {author}
  {\bibfnamefont {B.}~\bibnamefont {Ramshaw}}, \bibinfo {author} {\bibfnamefont
  {R.}~\bibnamefont {Liang}}, \bibinfo {author} {\bibfnamefont
  {D.}~\bibnamefont {Bonn}}, \bibinfo {author} {\bibfnamefont {W.}~\bibnamefont
  {Hardy}}, \bibinfo {author} {\bibfnamefont {N.}~\bibnamefont
  {Doiron-Leyraud}}, \bibinfo {author} {\bibfnamefont {A.}~\bibnamefont
  {Carrington}}, \bibinfo {author} {\bibfnamefont {N.}~\bibnamefont {Hussey}},
  \bibinfo {author} {\bibfnamefont {L.}~\bibnamefont {Taillefer}}, \ and\
  \bibinfo {author} {\bibfnamefont {C.}~\bibnamefont {Proust}},\ }\href@noop {}
  {\bibfield  {journal} {\bibinfo  {journal} {Comptes Rendus Physique}\
  }\textbf {\bibinfo {volume} {12}},\ \bibinfo {pages} {446} (\bibinfo {year}
  {2011})}\BibitemShut {NoStop}%
\bibitem [{\citenamefont {Sebastian}\ \emph {et~al.}(2012)\citenamefont
  {Sebastian}, \citenamefont {Harrison},\ and\ \citenamefont
  {Lonzarich}}]{sebastian_2012}%
  \BibitemOpen
  \bibfield  {author} {\bibinfo {author} {\bibfnamefont {S.~E.}\ \bibnamefont
  {Sebastian}}, \bibinfo {author} {\bibfnamefont {N.}~\bibnamefont {Harrison}},
  \ and\ \bibinfo {author} {\bibfnamefont {G.~G.}\ \bibnamefont {Lonzarich}},\
  }\href {\doibase 10.1088/0034-4885/75/10/102501} {\bibfield  {journal}
  {\bibinfo  {journal} {Rep. Prog. Phys.}\ }\textbf {\bibinfo {volume} {75}},\
  \bibinfo {pages} {102501} (\bibinfo {year} {2012})}\BibitemShut {NoStop}%
\bibitem [{\citenamefont {Wu}\ \emph {et~al.}(2011)\citenamefont {Wu},
  \citenamefont {Mayaffre}, \citenamefont {Kramer}, \citenamefont {Horvatic},
  \citenamefont {Berthier}, \citenamefont {Hardy}, \citenamefont {Liang},
  \citenamefont {Bonn},\ and\ \citenamefont {Julien}}]{wu_2011}%
  \BibitemOpen
  \bibfield  {author} {\bibinfo {author} {\bibfnamefont {T.}~\bibnamefont
  {Wu}}, \bibinfo {author} {\bibfnamefont {H.}~\bibnamefont {Mayaffre}},
  \bibinfo {author} {\bibfnamefont {S.}~\bibnamefont {Kramer}}, \bibinfo
  {author} {\bibfnamefont {M.}~\bibnamefont {Horvatic}}, \bibinfo {author}
  {\bibfnamefont {C.}~\bibnamefont {Berthier}}, \bibinfo {author}
  {\bibfnamefont {W.~N.}\ \bibnamefont {Hardy}}, \bibinfo {author}
  {\bibfnamefont {R.}~\bibnamefont {Liang}}, \bibinfo {author} {\bibfnamefont
  {D.~A.}\ \bibnamefont {Bonn}}, \ and\ \bibinfo {author} {\bibfnamefont
  {M.-H.}\ \bibnamefont {Julien}},\ }\href {\doibase 10.1038/nature10345}
  {\bibfield  {journal} {\bibinfo  {journal} {Nature}\ }\textbf {\bibinfo
  {volume} {477}},\ \bibinfo {pages} {191} (\bibinfo {year}
  {2011})}\BibitemShut {NoStop}%
\bibitem [{\citenamefont {Ghiringhelli}\ \emph {et~al.}(2012)\citenamefont
  {Ghiringhelli}, \citenamefont {Tacon}, \citenamefont {Minola}, \citenamefont
  {Blanco-Canosa}, \citenamefont {Mazzoli}, \citenamefont {Brookes},
  \citenamefont {Luca}, \citenamefont {Frano}, \citenamefont {Hawthorn},
  \citenamefont {He}, \citenamefont {Loew}, \citenamefont {Sala}, \citenamefont
  {Peets}, \citenamefont {Salluzzo}, \citenamefont {Schierle}, \citenamefont
  {Sutarto}, \citenamefont {Sawatzky}, \citenamefont {Weschke}, \citenamefont
  {Keimer},\ and\ \citenamefont {Braicovich}}]{ghiringhelli_2012}%
  \BibitemOpen
  \bibfield  {author} {\bibinfo {author} {\bibfnamefont {G.}~\bibnamefont
  {Ghiringhelli}}, \bibinfo {author} {\bibfnamefont {M.~L.}\ \bibnamefont
  {Tacon}}, \bibinfo {author} {\bibfnamefont {M.}~\bibnamefont {Minola}},
  \bibinfo {author} {\bibfnamefont {S.}~\bibnamefont {Blanco-Canosa}}, \bibinfo
  {author} {\bibfnamefont {C.}~\bibnamefont {Mazzoli}}, \bibinfo {author}
  {\bibfnamefont {N.~B.}\ \bibnamefont {Brookes}}, \bibinfo {author}
  {\bibfnamefont {G.~M.~D.}\ \bibnamefont {Luca}}, \bibinfo {author}
  {\bibfnamefont {A.}~\bibnamefont {Frano}}, \bibinfo {author} {\bibfnamefont
  {D.~G.}\ \bibnamefont {Hawthorn}}, \bibinfo {author} {\bibfnamefont
  {F.}~\bibnamefont {He}}, \bibinfo {author} {\bibfnamefont {T.}~\bibnamefont
  {Loew}}, \bibinfo {author} {\bibfnamefont {M.~M.}\ \bibnamefont {Sala}},
  \bibinfo {author} {\bibfnamefont {D.~C.}\ \bibnamefont {Peets}}, \bibinfo
  {author} {\bibfnamefont {M.}~\bibnamefont {Salluzzo}}, \bibinfo {author}
  {\bibfnamefont {E.}~\bibnamefont {Schierle}}, \bibinfo {author}
  {\bibfnamefont {R.}~\bibnamefont {Sutarto}}, \bibinfo {author} {\bibfnamefont
  {G.~A.}\ \bibnamefont {Sawatzky}}, \bibinfo {author} {\bibfnamefont
  {E.}~\bibnamefont {Weschke}}, \bibinfo {author} {\bibfnamefont
  {B.}~\bibnamefont {Keimer}}, \ and\ \bibinfo {author} {\bibfnamefont
  {L.}~\bibnamefont {Braicovich}},\ }\href {\doibase 10.1126/science.1223532}
  {\bibfield  {journal} {\bibinfo  {journal} {Science}\ }\textbf {\bibinfo
  {volume} {337}},\ \bibinfo {pages} {821} (\bibinfo {year}
  {2012})}\BibitemShut {NoStop}%
\bibitem [{\citenamefont {Chang}\ \emph {et~al.}(2012)\citenamefont {Chang},
  \citenamefont {Blackburn}, \citenamefont {Holmes}, \citenamefont
  {Christensen}, \citenamefont {Larsen}, \citenamefont {Mesot}, \citenamefont
  {Liang}, \citenamefont {Bonn}, \citenamefont {Hardy}, \citenamefont
  {Watenphul}, \citenamefont {Zimmermann}, \citenamefont {Forgan},\ and\
  \citenamefont {Hayden}}]{chang_2012}%
  \BibitemOpen
  \bibfield  {author} {\bibinfo {author} {\bibfnamefont {J.}~\bibnamefont
  {Chang}}, \bibinfo {author} {\bibfnamefont {E.}~\bibnamefont {Blackburn}},
  \bibinfo {author} {\bibfnamefont {A.~T.}\ \bibnamefont {Holmes}}, \bibinfo
  {author} {\bibfnamefont {N.~B.}\ \bibnamefont {Christensen}}, \bibinfo
  {author} {\bibfnamefont {J.}~\bibnamefont {Larsen}}, \bibinfo {author}
  {\bibfnamefont {J.}~\bibnamefont {Mesot}}, \bibinfo {author} {\bibfnamefont
  {R.}~\bibnamefont {Liang}}, \bibinfo {author} {\bibfnamefont {D.~A.}\
  \bibnamefont {Bonn}}, \bibinfo {author} {\bibfnamefont {W.~N.}\ \bibnamefont
  {Hardy}}, \bibinfo {author} {\bibfnamefont {A.}~\bibnamefont {Watenphul}},
  \bibinfo {author} {\bibfnamefont {M.~v.}\ \bibnamefont {Zimmermann}},
  \bibinfo {author} {\bibfnamefont {E.~M.}\ \bibnamefont {Forgan}}, \ and\
  \bibinfo {author} {\bibfnamefont {S.~M.}\ \bibnamefont {Hayden}},\ }\href
  {\doibase 10.1038/nphys2456} {\bibfield  {journal} {\bibinfo  {journal} {Nat.
  Phys.}\ }\textbf {\bibinfo {volume} {8}},\ \bibinfo {pages} {871} (\bibinfo
  {year} {2012})}\BibitemShut {NoStop}%
\bibitem [{\citenamefont {Wu}\ \emph {et~al.}(2013)\citenamefont {Wu},
  \citenamefont {Mayaffre}, \citenamefont {Krämer}, \citenamefont {Horvatic},
  \citenamefont {Berthier}, \citenamefont {Kuhns}, \citenamefont {Reyes},
  \citenamefont {Liang}, \citenamefont {Hardy}, \citenamefont {Bonn},\ and\
  \citenamefont {Julien}}]{wu_2013}%
  \BibitemOpen
  \bibfield  {author} {\bibinfo {author} {\bibfnamefont {T.}~\bibnamefont
  {Wu}}, \bibinfo {author} {\bibfnamefont {H.}~\bibnamefont {Mayaffre}},
  \bibinfo {author} {\bibfnamefont {S.}~\bibnamefont {Krämer}}, \bibinfo
  {author} {\bibfnamefont {M.}~\bibnamefont {Horvatic}}, \bibinfo {author}
  {\bibfnamefont {C.}~\bibnamefont {Berthier}}, \bibinfo {author}
  {\bibfnamefont {P.~L.}\ \bibnamefont {Kuhns}}, \bibinfo {author}
  {\bibfnamefont {A.~P.}\ \bibnamefont {Reyes}}, \bibinfo {author}
  {\bibfnamefont {R.}~\bibnamefont {Liang}}, \bibinfo {author} {\bibfnamefont
  {W.~N.}\ \bibnamefont {Hardy}}, \bibinfo {author} {\bibfnamefont {D.~A.}\
  \bibnamefont {Bonn}}, \ and\ \bibinfo {author} {\bibfnamefont {M.-H.}\
  \bibnamefont {Julien}},\ }\href@noop {} {\bibfield  {journal} {\bibinfo
  {journal} {Nat. Commun.}\ }\textbf {\bibinfo {volume} {4}} (\bibinfo {year}
  {2013})}\BibitemShut {NoStop}%
\bibitem [{\citenamefont {Li}\ \emph {et~al.}(2010)\citenamefont {Li},
  \citenamefont {Wang}, \citenamefont {Komiya}, \citenamefont {Ono},
  \citenamefont {Ando}, \citenamefont {Gu},\ and\ \citenamefont
  {Ong}}]{li_2010}%
  \BibitemOpen
  \bibfield  {author} {\bibinfo {author} {\bibfnamefont {L.}~\bibnamefont
  {Li}}, \bibinfo {author} {\bibfnamefont {Y.}~\bibnamefont {Wang}}, \bibinfo
  {author} {\bibfnamefont {S.}~\bibnamefont {Komiya}}, \bibinfo {author}
  {\bibfnamefont {S.}~\bibnamefont {Ono}}, \bibinfo {author} {\bibfnamefont
  {Y.}~\bibnamefont {Ando}}, \bibinfo {author} {\bibfnamefont {G.~D.}\
  \bibnamefont {Gu}}, \ and\ \bibinfo {author} {\bibfnamefont {N.~P.}\
  \bibnamefont {Ong}},\ }\href {\doibase 10.1103/PhysRevB.81.054510} {\bibfield
   {journal} {\bibinfo  {journal} {Phys. Rev. B}\ }\textbf {\bibinfo {volume}
  {81}},\ \bibinfo {pages} {054510} (\bibinfo {year} {2010})}\BibitemShut
  {NoStop}%
\bibitem [{\citenamefont {{LeBoeuf}}\ \emph {et~al.}(2007)\citenamefont
  {{LeBoeuf}}, \citenamefont {Doiron-Leyraud}, \citenamefont {Levallois},
  \citenamefont {Daou}, \citenamefont {Bonnemaison}, \citenamefont {Hussey},
  \citenamefont {Balicas}, \citenamefont {Ramshaw}, \citenamefont {Liang},
  \citenamefont {Bonn}, \citenamefont {Hardy}, \citenamefont {Adachi},
  \citenamefont {Proust},\ and\ \citenamefont {Taillefer}}]{leboeuf_2007}%
  \BibitemOpen
  \bibfield  {author} {\bibinfo {author} {\bibfnamefont {D.}~\bibnamefont
  {{LeBoeuf}}}, \bibinfo {author} {\bibfnamefont {N.}~\bibnamefont
  {Doiron-Leyraud}}, \bibinfo {author} {\bibfnamefont {J.}~\bibnamefont
  {Levallois}}, \bibinfo {author} {\bibfnamefont {R.}~\bibnamefont {Daou}},
  \bibinfo {author} {\bibfnamefont {J.-B.}\ \bibnamefont {Bonnemaison}},
  \bibinfo {author} {\bibfnamefont {N.~E.}\ \bibnamefont {Hussey}}, \bibinfo
  {author} {\bibfnamefont {L.}~\bibnamefont {Balicas}}, \bibinfo {author}
  {\bibfnamefont {B.~J.}\ \bibnamefont {Ramshaw}}, \bibinfo {author}
  {\bibfnamefont {R.}~\bibnamefont {Liang}}, \bibinfo {author} {\bibfnamefont
  {D.~A.}\ \bibnamefont {Bonn}}, \bibinfo {author} {\bibfnamefont {W.~N.}\
  \bibnamefont {Hardy}}, \bibinfo {author} {\bibfnamefont {S.}~\bibnamefont
  {Adachi}}, \bibinfo {author} {\bibfnamefont {C.}~\bibnamefont {Proust}}, \
  and\ \bibinfo {author} {\bibfnamefont {L.}~\bibnamefont {Taillefer}},\ }\href
  {\doibase 10.1038/nature06332} {\bibfield  {journal} {\bibinfo  {journal}
  {Nature}\ }\textbf {\bibinfo {volume} {450}},\ \bibinfo {pages} {533}
  (\bibinfo {year} {2007})}\BibitemShut {NoStop}%
\bibitem [{\citenamefont {{LeBoeuf}}\ \emph {et~al.}(2011)\citenamefont
  {{LeBoeuf}}, \citenamefont {Doiron-Leyraud}, \citenamefont {Vignolle},
  \citenamefont {Sutherland}, \citenamefont {Ramshaw}, \citenamefont
  {Levallois}, \citenamefont {Daou}, \citenamefont {Laliberte}, \citenamefont
  {Cyr-Choiniere}, \citenamefont {Chang}, \citenamefont {Jo}, \citenamefont
  {Balicas}, \citenamefont {Liang}, \citenamefont {Bonn}, \citenamefont
  {Hardy}, \citenamefont {Proust},\ and\ \citenamefont
  {Taillefer}}]{leboeuf_2011}%
  \BibitemOpen
  \bibfield  {author} {\bibinfo {author} {\bibfnamefont {D.}~\bibnamefont
  {{LeBoeuf}}}, \bibinfo {author} {\bibfnamefont {N.}~\bibnamefont
  {Doiron-Leyraud}}, \bibinfo {author} {\bibfnamefont {B.}~\bibnamefont
  {Vignolle}}, \bibinfo {author} {\bibfnamefont {M.}~\bibnamefont
  {Sutherland}}, \bibinfo {author} {\bibfnamefont {B.~J.}\ \bibnamefont
  {Ramshaw}}, \bibinfo {author} {\bibfnamefont {J.}~\bibnamefont {Levallois}},
  \bibinfo {author} {\bibfnamefont {R.}~\bibnamefont {Daou}}, \bibinfo {author}
  {\bibfnamefont {F.}~\bibnamefont {Laliberte}}, \bibinfo {author}
  {\bibfnamefont {O.}~\bibnamefont {Cyr-Choiniere}}, \bibinfo {author}
  {\bibfnamefont {J.}~\bibnamefont {Chang}}, \bibinfo {author} {\bibfnamefont
  {Y.~J.}\ \bibnamefont {Jo}}, \bibinfo {author} {\bibfnamefont
  {L.}~\bibnamefont {Balicas}}, \bibinfo {author} {\bibfnamefont
  {R.}~\bibnamefont {Liang}}, \bibinfo {author} {\bibfnamefont {D.~A.}\
  \bibnamefont {Bonn}}, \bibinfo {author} {\bibfnamefont {W.~N.}\ \bibnamefont
  {Hardy}}, \bibinfo {author} {\bibfnamefont {C.}~\bibnamefont {Proust}}, \
  and\ \bibinfo {author} {\bibfnamefont {L.}~\bibnamefont {Taillefer}},\ }\href
  {\doibase 10.1103/PhysRevB.83.054506} {\bibfield  {journal} {\bibinfo
  {journal} {Phys. Rev. B}\ }\textbf {\bibinfo {volume} {83}},\ \bibinfo
  {pages} {054506} (\bibinfo {year} {2011})}\BibitemShut {NoStop}%
\bibitem [{\citenamefont {Sebastian}\ \emph {et~al.}(2010)\citenamefont
  {Sebastian}, \citenamefont {Harrison}, \citenamefont {Altarawneh},
  \citenamefont {Liang}, \citenamefont {Bonn}, \citenamefont {Hardy},\ and\
  \citenamefont {Lonzarich}}]{sebastian_2010}%
  \BibitemOpen
  \bibfield  {author} {\bibinfo {author} {\bibfnamefont {S.~E.}\ \bibnamefont
  {Sebastian}}, \bibinfo {author} {\bibfnamefont {N.}~\bibnamefont {Harrison}},
  \bibinfo {author} {\bibfnamefont {M.~M.}\ \bibnamefont {Altarawneh}},
  \bibinfo {author} {\bibfnamefont {R.}~\bibnamefont {Liang}}, \bibinfo
  {author} {\bibfnamefont {D.~A.}\ \bibnamefont {Bonn}}, \bibinfo {author}
  {\bibfnamefont {W.~N.}\ \bibnamefont {Hardy}}, \ and\ \bibinfo {author}
  {\bibfnamefont {G.~G.}\ \bibnamefont {Lonzarich}},\ }\href {\doibase
  10.1103/PhysRevB.81.140505} {\bibfield  {journal} {\bibinfo  {journal} {Phys.
  Rev. B}\ }\textbf {\bibinfo {volume} {81}},\ \bibinfo {pages} {140505}
  (\bibinfo {year} {2010})}\BibitemShut {NoStop}%
\bibitem [{\citenamefont {Riggs}\ \emph {et~al.}(2011)\citenamefont {Riggs},
  \citenamefont {Vafek}, \citenamefont {Kemper}, \citenamefont {Betts},
  \citenamefont {Migliori}, \citenamefont {Balakirev}, \citenamefont {Hardy},
  \citenamefont {Liang}, \citenamefont {Bonn},\ and\ \citenamefont
  {Boebinger}}]{riggs_2011}%
  \BibitemOpen
  \bibfield  {author} {\bibinfo {author} {\bibfnamefont {S.~C.}\ \bibnamefont
  {Riggs}}, \bibinfo {author} {\bibfnamefont {O.}~\bibnamefont {Vafek}},
  \bibinfo {author} {\bibfnamefont {J.~B.}\ \bibnamefont {Kemper}}, \bibinfo
  {author} {\bibfnamefont {J.~B.}\ \bibnamefont {Betts}}, \bibinfo {author}
  {\bibfnamefont {A.}~\bibnamefont {Migliori}}, \bibinfo {author}
  {\bibfnamefont {F.~F.}\ \bibnamefont {Balakirev}}, \bibinfo {author}
  {\bibfnamefont {W.~N.}\ \bibnamefont {Hardy}}, \bibinfo {author}
  {\bibfnamefont {R.}~\bibnamefont {Liang}}, \bibinfo {author} {\bibfnamefont
  {D.~A.}\ \bibnamefont {Bonn}}, \ and\ \bibinfo {author} {\bibfnamefont
  {G.~S.}\ \bibnamefont {Boebinger}},\ }\href {\doibase 10.1038/nphys1921}
  {\bibfield  {journal} {\bibinfo  {journal} {Nat. Phys.}\ }\textbf {\bibinfo
  {volume} {7}},\ \bibinfo {pages} {332} (\bibinfo {year} {2011})}\BibitemShut
  {NoStop}%
\bibitem [{\citenamefont {Senthil}\ and\ \citenamefont
  {Lee}(2009)}]{senthil_2009}%
  \BibitemOpen
  \bibfield  {author} {\bibinfo {author} {\bibfnamefont {T.}~\bibnamefont
  {Senthil}}\ and\ \bibinfo {author} {\bibfnamefont {P.~A.}\ \bibnamefont
  {Lee}},\ }\href {\doibase 10.1103/PhysRevB.79.245116} {\bibfield  {journal}
  {\bibinfo  {journal} {Phys. Rev. B}\ }\textbf {\bibinfo {volume} {79}},\
  \bibinfo {pages} {245116} (\bibinfo {year} {2009})}\BibitemShut {NoStop}%
\bibitem [{\citenamefont {Banerjee}\ \emph {et~al.}(2013)\citenamefont
  {Banerjee}, \citenamefont {Zhang},\ and\ \citenamefont
  {Randeria}}]{banerjee_2013}%
  \BibitemOpen
  \bibfield  {author} {\bibinfo {author} {\bibfnamefont {S.}~\bibnamefont
  {Banerjee}}, \bibinfo {author} {\bibfnamefont {S.}~\bibnamefont {Zhang}}, \
  and\ \bibinfo {author} {\bibfnamefont {M.}~\bibnamefont {Randeria}},\ }\href
  {\doibase 10.1038/ncomms2667} {\bibfield  {journal} {\bibinfo  {journal}
  {Nat. Commun.}\ }\textbf {\bibinfo {volume} {4}},\ \bibinfo {pages} {1700}
  (\bibinfo {year} {2013})}\BibitemShut {NoStop}%
\bibitem [{\citenamefont {Scherpelz}\ \emph
  {et~al.}(2013{\natexlab{a}})\citenamefont {Scherpelz}, \citenamefont {Wulin},
  \citenamefont {Sopik}, \citenamefont {Levin},\ and\ \citenamefont
  {Rajagopal}}]{scherpelz_2013}%
  \BibitemOpen
  \bibfield  {author} {\bibinfo {author} {\bibfnamefont {P.}~\bibnamefont
  {Scherpelz}}, \bibinfo {author} {\bibfnamefont {D.}~\bibnamefont {Wulin}},
  \bibinfo {author} {\bibfnamefont {B.}~\bibnamefont {Sopik}}, \bibinfo
  {author} {\bibfnamefont {K.}~\bibnamefont {Levin}}, \ and\ \bibinfo {author}
  {\bibfnamefont {A.~K.}\ \bibnamefont {Rajagopal}},\ }\href {\doibase
  10.1103/PhysRevB.87.024516} {\bibfield  {journal} {\bibinfo  {journal} {Phys.
  Rev. B}\ }\textbf {\bibinfo {volume} {87}},\ \bibinfo {pages} {024516}
  (\bibinfo {year} {2013}{\natexlab{a}})}\BibitemShut {NoStop}%
\bibitem [{\citenamefont {Scherpelz}\ \emph
  {et~al.}(2013{\natexlab{b}})\citenamefont {Scherpelz}, \citenamefont {Wulin},
  \citenamefont {Levin},\ and\ \citenamefont {Rajagopal}}]{scherpelz_2013b}%
  \BibitemOpen
  \bibfield  {author} {\bibinfo {author} {\bibfnamefont {P.}~\bibnamefont
  {Scherpelz}}, \bibinfo {author} {\bibfnamefont {D.}~\bibnamefont {Wulin}},
  \bibinfo {author} {\bibfnamefont {K.}~\bibnamefont {Levin}}, \ and\ \bibinfo
  {author} {\bibfnamefont {A.~K.}\ \bibnamefont {Rajagopal}},\ }\href {\doibase
  10.1103/PhysRevA.87.063602} {\bibfield  {journal} {\bibinfo  {journal} {Phys.
  Rev. A}\ }\textbf {\bibinfo {volume} {87}},\ \bibinfo {pages} {063602}
  (\bibinfo {year} {2013}{\natexlab{b}})}\BibitemShut {NoStop}%
\bibitem [{\citenamefont {Janko}\ \emph {et~al.}(1997)\citenamefont {Janko},
  \citenamefont {Maly},\ and\ \citenamefont {Levin}}]{janko_1997}%
  \BibitemOpen
  \bibfield  {author} {\bibinfo {author} {\bibfnamefont {B.}~\bibnamefont
  {Janko}}, \bibinfo {author} {\bibfnamefont {J.}~\bibnamefont {Maly}}, \ and\
  \bibinfo {author} {\bibfnamefont {K.}~\bibnamefont {Levin}},\ }\href
  {\doibase 10.1103/PhysRevB.56.R11407} {\bibfield  {journal} {\bibinfo
  {journal} {Phys. Rev. B}\ }\textbf {\bibinfo {volume} {56}},\ \bibinfo
  {pages} {R11407} (\bibinfo {year} {1997})}\BibitemShut {NoStop}%
\bibitem [{\citenamefont {Maly}\ \emph {et~al.}(1997)\citenamefont {Maly},
  \citenamefont {Janko},\ and\ \citenamefont {Levin}}]{maly_1997b}%
  \BibitemOpen
  \bibfield  {author} {\bibinfo {author} {\bibfnamefont {J.}~\bibnamefont
  {Maly}}, \bibinfo {author} {\bibfnamefont {B.}~\bibnamefont {Janko}}, \ and\
  \bibinfo {author} {\bibfnamefont {K.}~\bibnamefont {Levin}},\ }\href@noop {}
  {\  (\bibinfo {year} {1997})}\BibitemShut {NoStop}%
\bibitem [{\citenamefont {Norman}\ \emph {et~al.}(1998)\citenamefont {Norman},
  \citenamefont {Randeria}, \citenamefont {Ding},\ and\ \citenamefont
  {Campuzano}}]{norman_1998}%
  \BibitemOpen
  \bibfield  {author} {\bibinfo {author} {\bibfnamefont {M.~R.}\ \bibnamefont
  {Norman}}, \bibinfo {author} {\bibfnamefont {M.}~\bibnamefont {Randeria}},
  \bibinfo {author} {\bibfnamefont {H.}~\bibnamefont {Ding}}, \ and\ \bibinfo
  {author} {\bibfnamefont {J.~C.}\ \bibnamefont {Campuzano}},\ }\href {\doibase
  10.1103/PhysRevB.57.R11093} {\bibfield  {journal} {\bibinfo  {journal} {Phys.
  Rev. B}\ }\textbf {\bibinfo {volume} {57}},\ \bibinfo {pages} {R11093}
  (\bibinfo {year} {1998})}\BibitemShut {NoStop}%
\bibitem [{\citenamefont {Norman}\ \emph {et~al.}(2007)\citenamefont {Norman},
  \citenamefont {Kanigel}, \citenamefont {Randeria}, \citenamefont
  {Chatterjee},\ and\ \citenamefont {Campuzano}}]{norman_2007}%
  \BibitemOpen
  \bibfield  {author} {\bibinfo {author} {\bibfnamefont {M.~R.}\ \bibnamefont
  {Norman}}, \bibinfo {author} {\bibfnamefont {A.}~\bibnamefont {Kanigel}},
  \bibinfo {author} {\bibfnamefont {M.}~\bibnamefont {Randeria}}, \bibinfo
  {author} {\bibfnamefont {U.}~\bibnamefont {Chatterjee}}, \ and\ \bibinfo
  {author} {\bibfnamefont {J.~C.}\ \bibnamefont {Campuzano}},\ }\href {\doibase
  10.1103/PhysRevB.76.174501} {\bibfield  {journal} {\bibinfo  {journal} {Phys.
  Rev. B}\ }\textbf {\bibinfo {volume} {76}},\ \bibinfo {pages} {174501}
  (\bibinfo {year} {2007})}\BibitemShut {NoStop}%
\bibitem [{\citenamefont {Dukan}\ and\ \citenamefont
  {Tesanovic}(1994)}]{dukan_1994}%
  \BibitemOpen
  \bibfield  {author} {\bibinfo {author} {\bibfnamefont {S.}~\bibnamefont
  {Dukan}}\ and\ \bibinfo {author} {\bibfnamefont {Z.}~\bibnamefont
  {Tesanovic}},\ }\href {\doibase 10.1103/PhysRevB.49.13017} {\bibfield
  {journal} {\bibinfo  {journal} {Phys. Rev. B}\ }\textbf {\bibinfo {volume}
  {49}},\ \bibinfo {pages} {13017} (\bibinfo {year} {1994})}\BibitemShut
  {NoStop}%
\bibitem [{Note1()}]{Note1}%
  \BibitemOpen
  \bibinfo {note} {Note that in $s$-wave gaps, nodal states are created solely
  by real-space inhomogeneity. However, in $d$-wave contributions to nodal
  states from pairing symmetry and Landau level-based real-space inhomogeneity
  are both present and become essentially inseparable,\cite {vavilov_1998}
  which means we will not distinguish between their descriptions
  here.}\BibitemShut {Stop}%
\bibitem [{\citenamefont {Vavilov}\ and\ \citenamefont
  {Mineev}(1998)}]{vavilov_1998}%
  \BibitemOpen
  \bibfield  {author} {\bibinfo {author} {\bibfnamefont {M.}~\bibnamefont
  {Vavilov}}\ and\ \bibinfo {author} {\bibfnamefont {V.}~\bibnamefont
  {Mineev}},\ }\href {\doibase 10.1134/1.558590} {\bibfield  {journal}
  {\bibinfo  {journal} {J. Exp. Theor. Phys.}\ }\textbf {\bibinfo {volume}
  {86}},\ \bibinfo {pages} {1191} (\bibinfo {year} {1998})}\BibitemShut
  {NoStop}%
\bibitem [{\citenamefont {Tesanovic}\ and\ \citenamefont
  {Sacramento}(1998)}]{tesanovic_1998}%
  \BibitemOpen
  \bibfield  {author} {\bibinfo {author} {\bibfnamefont {Z.}~\bibnamefont
  {Tesanovic}}\ and\ \bibinfo {author} {\bibfnamefont {P.~D.}\ \bibnamefont
  {Sacramento}},\ }\href {\doibase 10.1103/PhysRevLett.80.1521} {\bibfield
  {journal} {\bibinfo  {journal} {Phys. Rev. Lett.}\ }\textbf {\bibinfo
  {volume} {80}},\ \bibinfo {pages} {1521} (\bibinfo {year}
  {1998})}\BibitemShut {NoStop}%
\bibitem [{\citenamefont {Stephen}(1992)}]{stephen_1992}%
  \BibitemOpen
  \bibfield  {author} {\bibinfo {author} {\bibfnamefont {M.~J.}\ \bibnamefont
  {Stephen}},\ }\href {\doibase 10.1103/PhysRevB.45.5481} {\bibfield  {journal}
  {\bibinfo  {journal} {Phys. Rev. B}\ }\textbf {\bibinfo {volume} {45}},\
  \bibinfo {pages} {5481} (\bibinfo {year} {1992})}\BibitemShut {NoStop}%
\bibitem [{\citenamefont {Grissonnanche}\ \emph {et~al.}(2013)\citenamefont
  {Grissonnanche}, \citenamefont {Cyr-Choiniere}, \citenamefont {Laliberte},
  \citenamefont {de~Cotret}, \citenamefont {Juneau-Fecteau}, \citenamefont
  {Dufour-Beausejour}, \citenamefont {Delage}, \citenamefont {{LeBoeuf}},
  \citenamefont {Chang}, \citenamefont {Ramshaw}, \citenamefont {Bonn},
  \citenamefont {Hardy}, \citenamefont {Liang}, \citenamefont {Adachi},
  \citenamefont {Hussey}, \citenamefont {Vignolle}, \citenamefont {Proust},
  \citenamefont {Sutherland}, \citenamefont {Kramer}, \citenamefont {Park},
  \citenamefont {Graf}, \citenamefont {Doiron-Leyraud},\ and\ \citenamefont
  {Taillefer}}]{grissonnanche_2013}%
  \BibitemOpen
  \bibfield  {author} {\bibinfo {author} {\bibfnamefont {G.}~\bibnamefont
  {Grissonnanche}}, \bibinfo {author} {\bibfnamefont {O.}~\bibnamefont
  {Cyr-Choiniere}}, \bibinfo {author} {\bibfnamefont {F.}~\bibnamefont
  {Laliberte}}, \bibinfo {author} {\bibfnamefont {S.~R.}\ \bibnamefont
  {de~Cotret}}, \bibinfo {author} {\bibfnamefont {A.}~\bibnamefont
  {Juneau-Fecteau}}, \bibinfo {author} {\bibfnamefont {S.}~\bibnamefont
  {Dufour-Beausejour}}, \bibinfo {author} {\bibfnamefont {M.-E.}\ \bibnamefont
  {Delage}}, \bibinfo {author} {\bibfnamefont {D.}~\bibnamefont {{LeBoeuf}}},
  \bibinfo {author} {\bibfnamefont {J.}~\bibnamefont {Chang}}, \bibinfo
  {author} {\bibfnamefont {B.~J.}\ \bibnamefont {Ramshaw}}, \bibinfo {author}
  {\bibfnamefont {D.~A.}\ \bibnamefont {Bonn}}, \bibinfo {author}
  {\bibfnamefont {W.~N.}\ \bibnamefont {Hardy}}, \bibinfo {author}
  {\bibfnamefont {R.}~\bibnamefont {Liang}}, \bibinfo {author} {\bibfnamefont
  {S.}~\bibnamefont {Adachi}}, \bibinfo {author} {\bibfnamefont {N.~E.}\
  \bibnamefont {Hussey}}, \bibinfo {author} {\bibfnamefont {B.}~\bibnamefont
  {Vignolle}}, \bibinfo {author} {\bibfnamefont {C.}~\bibnamefont {Proust}},
  \bibinfo {author} {\bibfnamefont {M.}~\bibnamefont {Sutherland}}, \bibinfo
  {author} {\bibfnamefont {S.}~\bibnamefont {Kramer}}, \bibinfo {author}
  {\bibfnamefont {J.-H.}\ \bibnamefont {Park}}, \bibinfo {author}
  {\bibfnamefont {D.}~\bibnamefont {Graf}}, \bibinfo {author} {\bibfnamefont
  {N.}~\bibnamefont {Doiron-Leyraud}}, \ and\ \bibinfo {author} {\bibfnamefont
  {L.}~\bibnamefont {Taillefer}},\ }\href@noop {} {\  (\bibinfo {year}
  {2013})}\BibitemShut {NoStop}%
\bibitem [{\citenamefont {Tan}\ and\ \citenamefont {Levin}(2004)}]{tan_2004}%
  \BibitemOpen
  \bibfield  {author} {\bibinfo {author} {\bibfnamefont {S.}~\bibnamefont
  {Tan}}\ and\ \bibinfo {author} {\bibfnamefont {K.}~\bibnamefont {Levin}},\
  }\href@noop {} {\bibfield  {journal} {\bibinfo  {journal} {Phys. Rev. B}\
  }\textbf {\bibinfo {volume} {69}},\ \bibinfo {pages} {064510} (\bibinfo
  {year} {2004})}\BibitemShut {NoStop}%
\bibitem [{Note2()}]{Note2}%
  \BibitemOpen
  \bibinfo {note} {We take $\gamma /\Delta $, which would be zero in the
  superconducting state, to be as large as $0.5$ based on estimates which use
  the Fermi arc size in zero-field.\cite {norman_2007,he_2013} The choice of
  $\gamma '$ has little effect on the conclusions, but $\gamma ' = 0.2$ strikes
  a balance between computability (with no extremely sharp features, as opposed
  to $\gamma ' \rightarrow 0$) and the visibility of distinct oscillations (as
  opposed to large $\gamma '$). Parameters used for the model from
  Ref.~\protect \rev@citealpnum {senthil_2009} preserve that paper's choice of
  $\Gamma /\Delta $ and $\gamma /\Delta $, while $\Delta $ itself is scaled to
  match the other models.}\BibitemShut {Stop}%
\bibitem [{\citenamefont {He}\ \emph {et~al.}(2013)\citenamefont {He},
  \citenamefont {Scherpelz},\ and\ \citenamefont {Levin}}]{he_2013}%
  \BibitemOpen
  \bibfield  {author} {\bibinfo {author} {\bibfnamefont {Y.}~\bibnamefont
  {He}}, \bibinfo {author} {\bibfnamefont {P.}~\bibnamefont {Scherpelz}}, \
  and\ \bibinfo {author} {\bibfnamefont {K.}~\bibnamefont {Levin}},\ }\href
  {\doibase 10.1103/PhysRevB.88.064516} {\bibfield  {journal} {\bibinfo
  {journal} {Phys. Rev. B}\ }\textbf {\bibinfo {volume} {88}},\ \bibinfo
  {pages} {064516} (\bibinfo {year} {2013})}\BibitemShut {NoStop}%
\bibitem [{\citenamefont {Corcoran}\ \emph {et~al.}(1994)\citenamefont
  {Corcoran}, \citenamefont {Harrison}, \citenamefont {Hayden}, \citenamefont
  {Meeson}, \citenamefont {Springford},\ and\ \citenamefont {van~der
  Wel}}]{corcoran_1994}%
  \BibitemOpen
  \bibfield  {author} {\bibinfo {author} {\bibfnamefont {R.}~\bibnamefont
  {Corcoran}}, \bibinfo {author} {\bibfnamefont {N.}~\bibnamefont {Harrison}},
  \bibinfo {author} {\bibfnamefont {S.~M.}\ \bibnamefont {Hayden}}, \bibinfo
  {author} {\bibfnamefont {P.}~\bibnamefont {Meeson}}, \bibinfo {author}
  {\bibfnamefont {M.}~\bibnamefont {Springford}}, \ and\ \bibinfo {author}
  {\bibfnamefont {P.~J.}\ \bibnamefont {van~der Wel}},\ }\href {\doibase
  10.1103/PhysRevLett.72.701} {\bibfield  {journal} {\bibinfo  {journal} {Phys.
  Rev. Lett.}\ }\textbf {\bibinfo {volume} {72}},\ \bibinfo {pages} {701}
  (\bibinfo {year} {1994})}\BibitemShut {NoStop}%
\bibitem [{\citenamefont {Maniv}\ \emph {et~al.}(1992)\citenamefont {Maniv},
  \citenamefont {Rom}, \citenamefont {Vagner},\ and\ \citenamefont
  {Wyder}}]{maniv_1992}%
  \BibitemOpen
  \bibfield  {author} {\bibinfo {author} {\bibfnamefont {T.}~\bibnamefont
  {Maniv}}, \bibinfo {author} {\bibfnamefont {A.~I.}\ \bibnamefont {Rom}},
  \bibinfo {author} {\bibfnamefont {I.~D.}\ \bibnamefont {Vagner}}, \ and\
  \bibinfo {author} {\bibfnamefont {P.}~\bibnamefont {Wyder}},\ }\href
  {\doibase 10.1103/PhysRevB.46.8360} {\bibfield  {journal} {\bibinfo
  {journal} {Phys. Rev. B}\ }\textbf {\bibinfo {volume} {46}},\ \bibinfo
  {pages} {8360} (\bibinfo {year} {1992})}\BibitemShut {NoStop}%
\bibitem [{\citenamefont {Norman}\ \emph {et~al.}(1995)\citenamefont {Norman},
  \citenamefont {{MacDonald}},\ and\ \citenamefont {Akera}}]{norman_1995}%
  \BibitemOpen
  \bibfield  {author} {\bibinfo {author} {\bibfnamefont {M.~R.}\ \bibnamefont
  {Norman}}, \bibinfo {author} {\bibfnamefont {A.~H.}\ \bibnamefont
  {{MacDonald}}}, \ and\ \bibinfo {author} {\bibfnamefont {H.}~\bibnamefont
  {Akera}},\ }\href {\doibase 10.1103/PhysRevB.51.5927} {\bibfield  {journal}
  {\bibinfo  {journal} {Phys. Rev. B}\ }\textbf {\bibinfo {volume} {51}},\
  \bibinfo {pages} {5927} (\bibinfo {year} {1995})}\BibitemShut {NoStop}%
\bibitem [{\citenamefont {Janssen}\ \emph {et~al.}(1998)\citenamefont
  {Janssen}, \citenamefont {Haworth}, \citenamefont {Hayden}, \citenamefont
  {Meeson}, \citenamefont {Springford},\ and\ \citenamefont
  {Wasserman}}]{janssen_1998}%
  \BibitemOpen
  \bibfield  {author} {\bibinfo {author} {\bibfnamefont {T.~J. B.~M.}\
  \bibnamefont {Janssen}}, \bibinfo {author} {\bibfnamefont {C.}~\bibnamefont
  {Haworth}}, \bibinfo {author} {\bibfnamefont {S.~M.}\ \bibnamefont {Hayden}},
  \bibinfo {author} {\bibfnamefont {P.}~\bibnamefont {Meeson}}, \bibinfo
  {author} {\bibfnamefont {M.}~\bibnamefont {Springford}}, \ and\ \bibinfo
  {author} {\bibfnamefont {A.}~\bibnamefont {Wasserman}},\ }\href {\doibase
  10.1103/PhysRevB.57.11698} {\bibfield  {journal} {\bibinfo  {journal} {Phys.
  Rev. B}\ }\textbf {\bibinfo {volume} {57}},\ \bibinfo {pages} {11698}
  (\bibinfo {year} {1998})}\BibitemShut {NoStop}%
\bibitem [{\citenamefont {Dukan}\ and\ \citenamefont
  {Tesanovic}(1995)}]{dukan_1995}%
  \BibitemOpen
  \bibfield  {author} {\bibinfo {author} {\bibfnamefont {S.}~\bibnamefont
  {Dukan}}\ and\ \bibinfo {author} {\bibfnamefont {Z.}~\bibnamefont
  {Tesanovic}},\ }\href {\doibase 10.1103/PhysRevLett.74.2311} {\bibfield
  {journal} {\bibinfo  {journal} {Phys. Rev. Lett.}\ }\textbf {\bibinfo
  {volume} {74}},\ \bibinfo {pages} {2311} (\bibinfo {year}
  {1995})}\BibitemShut {NoStop}%
\bibitem [{\citenamefont {Shibauchi}\ \emph {et~al.}(2001)\citenamefont
  {Shibauchi}, \citenamefont {Krusin-Elbaum}, \citenamefont {Li}, \citenamefont
  {Maley},\ and\ \citenamefont {Kes}}]{shibauchi_2001}%
  \BibitemOpen
  \bibfield  {author} {\bibinfo {author} {\bibfnamefont {T.}~\bibnamefont
  {Shibauchi}}, \bibinfo {author} {\bibfnamefont {L.}~\bibnamefont
  {Krusin-Elbaum}}, \bibinfo {author} {\bibfnamefont {M.}~\bibnamefont {Li}},
  \bibinfo {author} {\bibfnamefont {M.~P.}\ \bibnamefont {Maley}}, \ and\
  \bibinfo {author} {\bibfnamefont {P.~H.}\ \bibnamefont {Kes}},\ }\href
  {\doibase 10.1103/PhysRevLett.86.5763} {\bibfield  {journal} {\bibinfo
  {journal} {Phys. Rev. Lett.}\ }\textbf {\bibinfo {volume} {86}},\ \bibinfo
  {pages} {5763} (\bibinfo {year} {2001})}\BibitemShut {NoStop}%
\bibitem [{\citenamefont {Fetter}\ and\ \citenamefont {~}(1971)}]{fetter_1971}%
  \BibitemOpen
  \bibfield  {author} {\bibinfo {author} {\bibfnamefont {A.~L.}\ \bibnamefont
  {Fetter}}\ and\ \bibinfo {author} {\bibfnamefont {J.~D.}\ \bibnamefont {~},
  \bibfnamefont {Walecka}},\ }\href@noop {} {\emph {\bibinfo {title} {Quantum
  theory of many-particle systems}}}\ (\bibinfo  {publisher} {{McGraw-Hill}},\
  \bibinfo {address} {San Francisco},\ \bibinfo {year} {1971})\BibitemShut
  {NoStop}%
\bibitem [{\citenamefont {Abrikosov}(1988)}]{abrikosov_1988}%
  \BibitemOpen
  \bibfield  {author} {\bibinfo {author} {\bibfnamefont {A.}~\bibnamefont
  {Abrikosov}},\ }\href@noop {} {\emph {\bibinfo {title} {Fundamentals of the
  Theory of Metals}}}\ (\bibinfo  {publisher} {North Holland},\ \bibinfo {year}
  {1988})\BibitemShut {NoStop}%
\bibitem [{\citenamefont {Shoenberg}(1984)}]{shoenberg_1984}%
  \BibitemOpen
  \bibfield  {author} {\bibinfo {author} {\bibfnamefont {D.}~\bibnamefont
  {Shoenberg}},\ }\href@noop {} {\emph {\bibinfo {title} {Magnetic Oscillations
  in Metals}}}\ (\bibinfo  {publisher} {Cambridge University Press},\ \bibinfo
  {year} {1984})\BibitemShut {NoStop}%
\bibitem [{Note3()}]{Note3}%
  \BibitemOpen
  \bibinfo {note} {For our calculation, we make one more transformation, by
  splitting up the $\omega $ integral in Eq.~\protect \textup {\hbox
  {\mathsurround \z@ \protect \normalfont (\ignorespaces \ref
  {EqFinalTD}\unskip \@@italiccorr )}} (with $I$ the integrand): $\DOTSI \intop
  \ilimits@ _{-\infty }^y d\omega \ I(\omega ) \rightarrow \DOTSI \intop
  \ilimits@ _0^y d\omega \ I(\omega ) + \DOTSI \intop \ilimits@ _{-\infty }^0
  d\omega \ I(\omega ).$ We know that as $T \rightarrow \infty $, the result of
  Eq.~\protect \textup {\hbox {\mathsurround \z@ \protect \normalfont
  (\ignorespaces \ref {EqFinalTD}\unskip \@@italiccorr )}} must go to zero, and
  with $\DOTSI \intop \ilimits@ _{-\infty }^\infty dy\ (df(y)/dy) = 1$, we find
  that $\DOTSI \intop \ilimits@ _{-\infty }^0 d\omega \ I(\omega ) = -\protect
  \qopname \relax m{lim}_{T \rightarrow \infty } \DOTSI \intop \ilimits@
  _{-\infty }^\infty dy\ (df(y)/dy) \DOTSI \intop \ilimits@ _{0}^y d\omega \
  I(\omega )$. This latter formula is used to compute the constant part of the
  original integral, so that each individual calculation only requires
  computation of the $\DOTSI \intop \ilimits@ _0^y d\omega \ I(\omega )$
  term.}\BibitemShut {Stop}%
\bibitem [{Note4()}]{Note4}%
  \BibitemOpen
  \bibinfo {note} {Note that unlike Eq.~(1) in Ref.~\protect \rev@citealpnum
  {sebastian_2010}, no $\eta _{\protect \mathrm {lim}}$ is used
  here.}\BibitemShut {Stop}%
\end{thebibliography}%

\end{document}